\documentclass[reprint,superscriptaddress,amsmath,amssymb,pre]{revtex4-1}

\usepackage{graphicx}
\usepackage{dcolumn}
\usepackage{bm}
\usepackage{amssymb}
\usepackage{color,soul}
\usepackage{hyperref}

\newcommand{\includegraphic}[5][,]{%
	\setbox1=\hbox{\includegraphics[#1]{#2}}
	\leavevmode\rlap{\usebox1}
	\rlap{\hspace*{#4}\raisebox{\dimexpr\ht1-#5\baselineskip}{\normalsize{#3}}}
	\phantom{\usebox1}}

\begin{document}
	
\preprint{APS/123-QED}
	
\title{Coupling conditions for globally stable and robust synchrony of chaotic systems }
	
	\author{Suman Saha}
	\email{ecesuman06@gmail.com}
\affiliation{Department of Instrumentation and Electronics Engineering, Jadavpur University, Kolkata 700090, India} 
\affiliation{Department of Electronics, Asutosh College, Kolkata 700026, India}
\affiliation{Dumkal Institute of Engineering and Technology,  Murshidabad 742406, India}
\author{Arindam Mishra}
\email{arindammishra@gmail.com}
\affiliation{Department of Physics, Jadavpur University, Kolkata 700032, India}
\author{E. Padmanaban}
\affiliation{CSIR-Indian Institute of Chemical Biology, Kolkata 700032, India}
\affiliation{Center for Complex System Research Kolkata, Kolkata 700094, India}		
\author{Sourav K. Bhowmick}
\affiliation{Department of Electronics, Asutosh College, Kolkata 700026, India}
\affiliation{Center for Complex System Research Kolkata, Kolkata 700094, India}
\author{Prodyot K. Roy}
\affiliation{Center for Complex System Research Kolkata, Kolkata 700094, India}
\affiliation{Department of Mathematics, Presidency University, Kolkata 700073, India}
\author{Bivas Dam}
\affiliation{Department of Instrumentation and Electronics Engineering, Jadavpur University, Kolkata 700090, India} 
\author{Syamal K. Dana}
\email{syamaldana@gmail.com}
\affiliation{Center for Complex System Research Kolkata, Kolkata 700094, India}
\affiliation{Department of Mathematics, Jadavpur University, Kolkata 700032, India}

	\date{\today}
		
	\begin{abstract}
 We propose a   set of general coupling conditions to select a coupling profile (a set of coupling matrices) from the linear flow matrix (LFM) of dynamical systems for realizing global stability of complete synchronization (CS) in identical systems and robustness to parameter perturbation. The coupling matrices   define the coupling links  between any two oscillators in a network that consists of a conventional diffusive coupling link (self-coupling link) as well as a cross-coupling link. The addition of a selective cross-coupling link in particular plays constructive roles that ensure the global stability of synchrony and furthermore enables robustness of synchrony against  small to non-small parameter perturbation.  We  elaborate the general conditions for the selection of coupling profiles for  two coupled systems,  three- and four-node network motifs analytically as well as numerically using benchmark models, the Lorenz system, the Hindmarsh-Rose neuron model, the Shimizu-Morioka laser model, the R\"ossler system and  a Sprott system. The role of  the cross-coupling link is, particularly, exemplified with an example of a larger network where it  saves the network from a breakdown of synchrony against large parameter perturbation in any node. The perturbed node  in the network transits from CS to generalized synchronization (GS) when all the other nodes remain in CS. The GS is manifested  by an amplified response of the perturbed node in a coherent state.  
	\end{abstract}
	
	\maketitle	
	\section{Introduction}
 Synchrony is a most desirable state in many real-world dynamical networks such as the human brain \cite{Bablo, Lytt}, power grid \cite{Macho}, ecological network \cite{Hirota},  
 which must be stable against small to non-small perturbations.  
 An instability due to a perturbation (intrinsic or extrinsic) may push a synchronous network  to the point  of   failure of a desired performance (schizophrenia, power black-out, desertification etc.). To derive stability of such networks' synchrony, the  master stability function (MSF) is a breakthrough concept \cite{Pecora}  that  determines the stability condition of  synchronization in networks of identical oscillators. It encompasses the role of a network structure and the dynamics of the nodes as well, however, synchrony is locally stable since it is based on linear stability analysis.  The MSF  stability   condition provided no clue how to prevent a   breakdown of synchronization or failure of the networks' desired function against a parameter perturbation. Recently, a basin stability approach \cite{Wiley, Menck} was undertaken to assess the stability region of the basin of attraction of a synchronous state in complex networks and to test  how rewiring of coupling links can change the degree of stability to a smaller or a larger subbasin.
 A global stability can indeed enlarge the synchronous basin of a dynamical network beyond local stability and provide robustness against small or non-small perturbation and, thereby  save a synchronous network from drifting away to a disaster. 
However, imposing global stability  of synchronization in a  dynamical network is still an enigma.
\par  Realization of global stability of a synchronized state in a relatively small ensemble of oscillators    is well known by a design of coupling \cite{Pad, Chen,Jiang} that uses the Lyapunov function stability (LFS) approach. 
 The main weakness of the design of coupling approach \cite{Pad, Chen,Jiang, Grosu} 
is the cost of coupling due to a relatively large number of coupling links  necessary for global stability of synchronization and the essential presence of nonlinear coupling. In the case of a parameter mismatch, the number of coupling links is even larger.
Earlier studies \cite{Motter,Gu,Hagberg,Duan} on dynamical networks have already shown that many coupling links does not mean better synchrony. By this time, it is also known that synchronization of dynamical networks can be enhanced by rewiring directed links \cite{Zeng1} or by deletion of redundant links \cite{Zeng2}.  
Playing \cite{Olusola} with the number of coupling links for finding  an appropriate choice  \cite{skardal} or  redundancy of links in a network \cite{Schult} is a judicious approach to improve the quality of synchronization.  A natural question arises if addition or deletion of suitable coupling links can enlarge the basin of a synchronous state and thereby save an ensemble of oscillators from the brink of a breakdown?  Is it possible to choose  an appropriate set of coupling links for an ensemble of oscillators that may ensure globally stable synchrony?  In a real world network, a parameter  perturbation (system or coupling parameter) may originate due to internal/external disturbance when synchrony breaks down. Establising robustness of a desired synchronous state against such parameter perturbation is an important issue to resolve although not an easy task.  
 We make an attempt here to address the issues by selecting an appropriate  coupling profile consisting of linear coupling only for any two nodes in a network of oscillators. 
\par   From our past studies \cite{Suman} of two coupled systems,  we know   that addition of cross-coupling links over and above the conventional diffusive  coupling 
can realize global stability of synchrony and adds robustness to induced heterogeneity. However, the choice of coupling links was arbitrarily decided by the LFS condition and there were several  available options. This leads us to a search for an appropriate coupling profile for dynamical systems, in general, with fewer  diffusive scalar coupling and seletive cross-coupling  that could suffice  a global stability of synchrony in an ensemble of the dynamical system. Another important target is to ensure   robustness of synchrony against heterogeneity due  to a parameter mismatch ranging from a small to a large value.
\par In this paper, we propose  a  set of general coupling conditions that allows a selection of  self-coupling and cross-coupling  matrices, which is  called as the coupling profile, from the LFM of  dynamical systems and  ensures global stability of a synchronous state in  a network of the  dynamical system. 
A coupling matrix determines the coupling functions that involve specific state variables of two  dynamical systems to build  up coupling links. Each coupling function is considered as a link between any two dynamical nodes. The conventional diffusive scalar coupling  is redefined here as self-coupling because it involves two similar variables of the coupled systems and added to the dynamics of the same state variables. The cross-coupling also involves two similar variables \cite{Suman, Mishra, Motter1} of the systems, but adds to the dynamics of a different state variable. 
 An appropriate selection of the cross-coupling link as well as the conventional self-coupling link realizes  global stability of CS, expands the range of critical coupling for CS beyond the range defined by the MSF and thereby adds robustness to any drifting of the coupling paramater and, furthermore achieves robustness against  perturbation in system parameters. As a result, if  any one of the coupled systems is manually detuned by a small or a large value, CS simply transits to a type of GS \cite{Rulkov}. 
 The detuned system's response is an amplified/attenuated replica of the unperturbed system. The  coupled system maintains the dynamics of the isolated system,  which we explain later in the text.
The salient features of  the selection of a coupling profile, especially, the addition of selective cross-coupling links, are exemplified when a larger network is considered. In  a large  network of dynamical systems, in a CS state, if a parameter of a node is perturbed that particular node transits to a  GS state when all the other nodes remain in the CS state. The perturbed node's attractor is an amplified replica of all the other nodes' attractors. 
\par We first presented an example of two-coupled Lorenz systems \cite{m-1} in section II to elaborate how addition of a cross-coupling link over and above the self-coupling realized global stability of synchrony and adds robustness to synchrony. Then we extended the results to frame a set of general conditions how a coupling profile can be  appropriately selected from the LFM of a dynamical system and validate the scheme, in section III, analytically and numerically, using the paradigmatic model systems (Shimizu-Morioka (SM) laser model \cite{m-3}, R\"ossler system \cite{m-3} and a Sprott system \cite{m-4}) for 2-coupled model systems. In section IV, we presented analytical details for a second example of a 2-coupled  Hindmarsh-Rose (HR)  model \cite{m-2}.  We succesfully extended the results to 3-, 4-node network motifs \cite{Alon} in section V.
Analytical details for 2-coupled systems were presented in an Appendix. For analytical details  of the network motifs, see the  Supplementary Material (SM). Finally,  an example of  a  larger network was presented in section VI with a summary in section VII.  
 \section{Coupled Lorenz systems}
\par
\begin{figure}[!ht]
	\hspace{-15pt}
 \includegraphic[width=4cm,height=3.5cm]{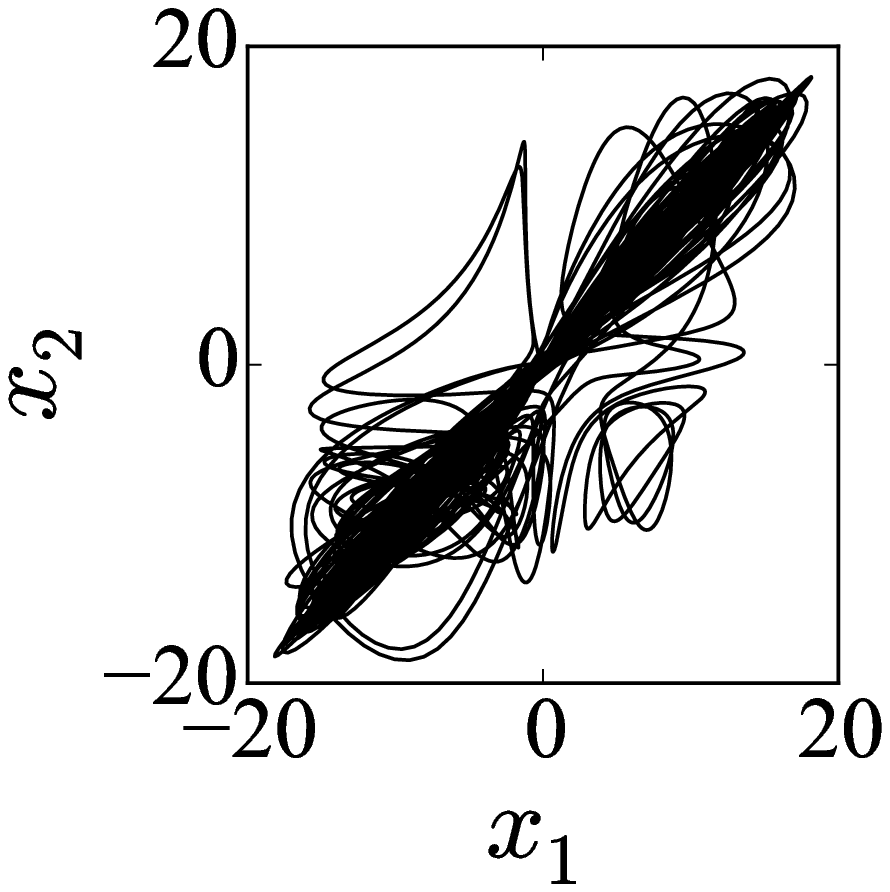}{(a)}{0pt}{0}
	\vspace{0pt}
	\includegraphic[width=4cm,height=3.5cm]{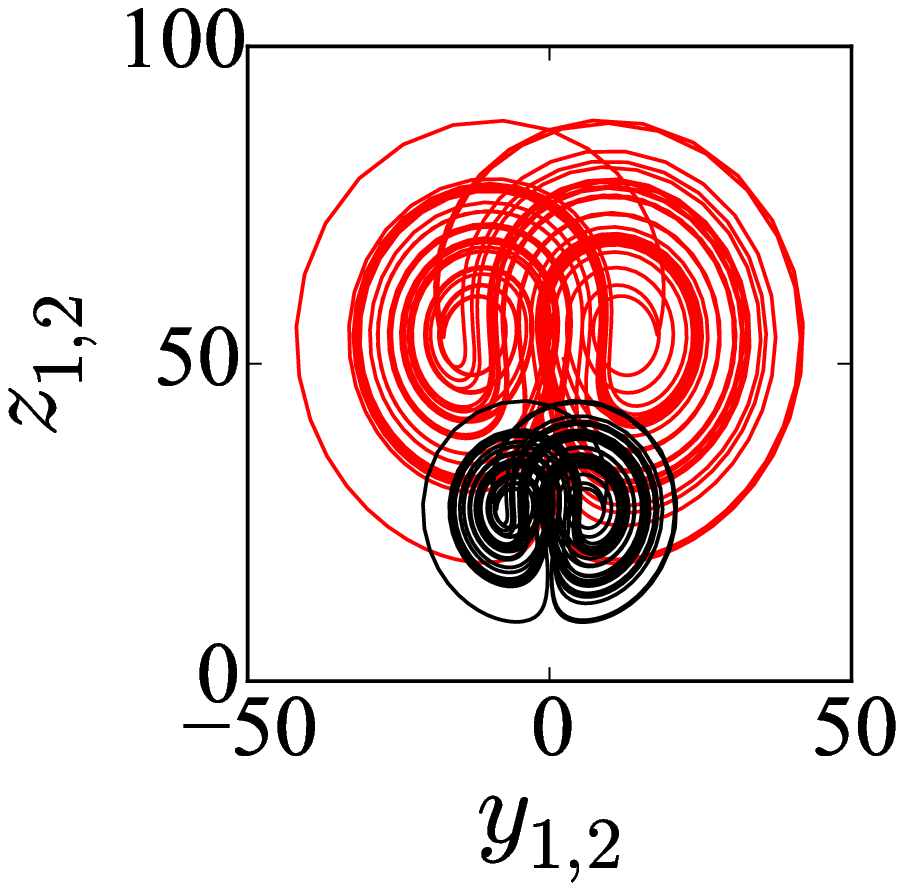}{(c)}{0pt}{0}\\
	\hspace{-15pt}
\includegraphic[width=4cm,height=3.5cm]{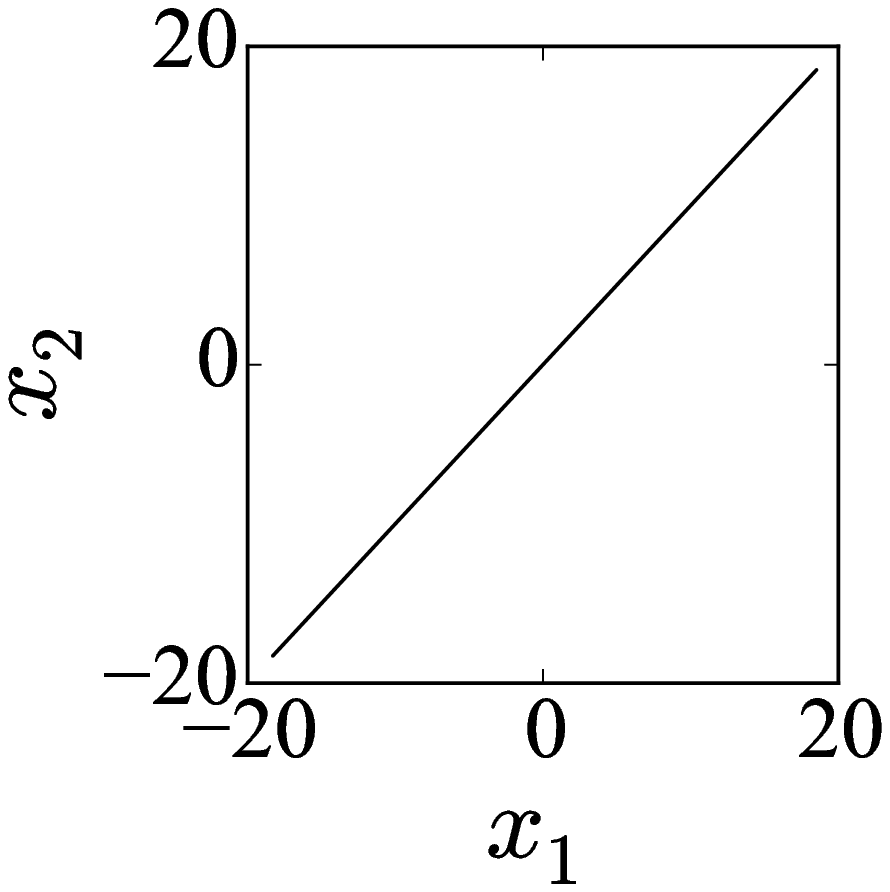}{(b)}{0pt}{0}
\includegraphic[width=4cm,height=3.5cm]{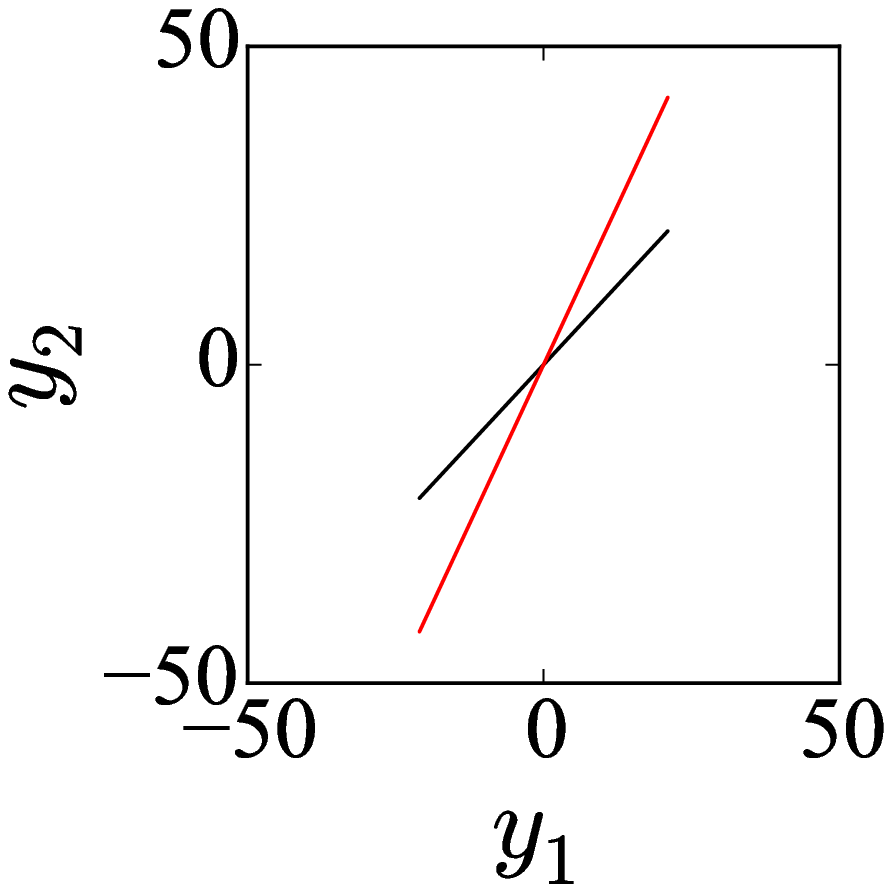}{(d)}{0pt}{0}
\caption{(Color online) Coupled Lorenz system. [$r_1$=$r_2$=28, $\sigma$=10, $b$=8/3].  $x_2$ vs $x_1$ plots  show (a) loss of synchrony  for a drifting $\epsilon_1=3.8<\epsilon_{MSF}$, (b) synchrony restored  by addition of one directed cross-coupling link ($\epsilon_2$=10), (c) an amplified attractor (red) along both $y$- and $z$-direction for detuning ($r_2$=56) and the reference system's attractor (black), (d) synchronization manifold $y_2$ vs. $y_1$: identical coupled oscillators in black and detuned oscillators in red. }\label{fig1}
\end{figure}

 
 Two identical Lorenz oscillators  are assumed synchronized under  a bidirectional  self-coupling, 
\begin{equation}
\begin{array}{l}
\dot{x}_{1,2}=\sigma({y}_{1,2}-{x}_{1,2}) + \epsilon_1(x_{2,1}-x_{1,2}) \\
\dot{y}_{1,2}=r_{1,2} x_{1,2}-y_{1,2}-x_{1,2}z_{1,2} \\
\dot{z}_{1,2}=-bz_{1,2}+x_{1,2} y_{1,2}
\end{array} \label{lor}
\end{equation}
where subscripts denote oscillators (1, 2). Two systems are assumed identical, $r_1=r_2=r$, and a critical coupling $\epsilon_1=\epsilon_{MSF}>3.9225$ is determined using the MSF when the CS state,  
$x_1=x_2$, $y_1=y_2$, $z_1=z_2$, emerges as locally stable. 
If the coupling strength drifts to a value lower than the critical value, $\epsilon_1=3.8 <\epsilon_{MSF}$, synchrony is lost (Fig.~\ref{fig1}a).
To mitigate this breakdown of synchrony, one directed cross-coupling link is added \cite{Pad} to the coupled system in Eq.~(1):  a linear diffusive coupling function involving the $y_{1,2}$-variables  is added to the dynamics of the $x_2$-variable of the second oscillator, 
\begin{equation}
\begin{array}{l}
\dot{x}_1 =\sigma({y}_{1}-{x}_{1}) + \epsilon_1(x_2-x_1) ,\\ 
\dot{x}_{2}=\sigma({y}_{2}
-{x}_{2}) +\epsilon_1(x_1-x_2)+\epsilon_2(y_1 -y_2) \\
\dot{y}_{1,2}=r_{1,2} x_{1,2}-y_{1,2}-x_{1,2}z_{1,2} \\
\dot{z}_{1,2}=-bz_{1,2}+x_{1,2} y_{1,2}
\end{array} \label{lor1} 
\end{equation}
where $\epsilon_2$ is the strength of the cross-coupling link. A directed cross-coupling link  is basically added to the second oscillator when the first oscillator's dynamics remains unchanged. The error dynamics  $\dot{e}_x=(\dot{x}_1-\dot{x}_2$) changes, 
\begin{equation}
\dot{e_x}
=\sigma({e_y}-{e_x}) - 2\epsilon_1e_x -\epsilon_2e_y, \label{ex3}
\end{equation}
while  $\dot e_y=\dot y_1-\dot y _2$, $\dot e_z=\dot z_1-\dot z_2$ remain unchanged. For a choice of Lyapunov function, $V(e_x) =\frac{1}{2}e_x^{2}$,
\begin{equation}
\dot{V}(e_x)= -(\sigma+2\epsilon_1)e_x^{2} + (\sigma-\epsilon_2)e_xe_y \label{Vex4}
\end{equation}
that establishes an asymptotic  stability of $x_1=x_2$ if  $\epsilon_1>-\sigma/2$ and $\epsilon_2=\sigma$. 
It leads to a stability relation of a Lyapunov function, $V'(e_y, e_z)=1/2e_y^2 +1/2e_z^2$, 
\begin{equation}
\dot{V'}(e_y,e_z)= -e_y^{2} -be_z^{2} <0 \label{Veyz5}
\end{equation}
that ensures global stability of  the CS manifold, $x_1=x_2$, $y_1=y_2$ and $z_1=z_2$, for identical systems. 
The addition of the directed cross-coupling link  expands the range of critical  self-coupling strength largely as defined by the condition, $\epsilon_1>-\sigma/2$, which is much lower than $\epsilon_{MSF}$. The global stability of CS  thus sustains (Fig.~\ref{fig1}b) a large drifting of the self-coupling strength below the critical value $\epsilon_{MSF}$. The synchronous dynamics maintains the isolated dynamics of the coupled systems since both the self- and cross-coupling functions are effectively vanished in the CS state (see Eq.\eqref{lor1}). 
\par Furthermore, the cross-coupling link  plays the most important role of realizing robustness of synchrony  against a system parameter perturbation. 
For a demonstration, a parameter  $r_2=(r_1+\Delta r$) is detuned manually in Eq.\eqref{lor1} when the  error dynamics  is modified (details are given in refs.~\cite{equ1,equ2})
$\dot{e}^*=\left[-(\sigma+2\epsilon_1)e_x+(\sigma-\epsilon_2)e_y,\; -e_y^*-x_1e_z^*, \; x_1e_y^*-be_z^*\right]^T$ 
where $e_y^*=\frac{r_2}{r_1}y_1-y_2$ and $e_z^*=\frac{r_2}{r_1}z_1-z_2$ and this leads to a revised stability condition of a modified Lyapunov function ($V^{*}$),
\begin{equation}
\begin{array}{l}
\dot{V}^{*}(e^*)=-\sigma e_x^2 -e_y^{*2}-be_z^{*2}<0.  \label{le4}
\end{array}
\end{equation}
provided $\epsilon_1>-\sigma/2$ and $\epsilon_2=\sigma$. A new globally stable  synchronous state, $x_2=x_1$,  $y_2=\frac{r_2}{r_1}y_1$ and $z_2=\frac{r_2}{r_1}z_1$ emerges which we  call as a type of GS state.  A linear functional relation emerges  between the coupled systems. As an example,  for  a detuning of $r_2$  by twice ($\frac{r_2}{r_1}$=2), the  new emergent coherent state,  $x_2=x_1$, $y_2=2y_1$ and $z_2=2z_1$, is globally stable. Effectively, the detuned system's attractor  (red) is amplified along the $y$- and $z$-direction exactly twice that of the unperturbed system's attractor (black) as shown in Fig.~\ref{fig1}(c). 
\par  CS simply transits to GS due to the parameter detuning without a loss of overall coherence or synchrony in the coupled system. This is  manifested by  a rotation  of the CS manifold ($y_1=y_2$ in black line) in Fig.~\ref{fig1}(d) to GS manifold ($y_2=2y_1$, red line). In the GS state, $x_1=x_2$ condition is still preserved when the self-coupling function vanishes. The first oscillator is thus not affected by the  cross-coupling and hence maintains the isolated dynamics and,  plays an effective role of a driver. The second oscillator also continues with the original dynamics because it simply works as a slave to the first oscillator in a coherent state, but with an amplified replica (red) as shown in Fig.~\ref{fig1}(c). If $r_1$ is detuned, instead, the qualitative behavior remains same except that the coupled dynamics will now change since the first oscillator has a change of dynamics.   The final effect is  a dramatic improvement in the quality of synchronization that is simply implemented by a selection of two linear coupling links, especially, the  addition of one selective  cross-coupling link. 
\section{Linear Flow Matrix: Coupling profile}
\par  
The particular selection of a coupling profile (self-coupling and cross-coupling links) for the Lorenz system, can really be made in a systematic manner from the LFM of the system and,  we can  frame a set of general coupling conditions for the selection of a coupling profile  for many dynamical systems,
\begin{equation}
\begin{array}{c}
\dot{\bf{x}}=g({\bf{x}})=F{\bf{x}} + L({\bf{x}}) + {\bf C}
\end{array}\label{lmatlor1}
\end{equation}
$g$:$\mathbb {R}^n\rightarrow \mathbb{R}^n$ is the flow of the system,  $F$ is the LFM (${n\times n}$ matrix, $n$=3 for our example systems), $\textbf{x}=[x,y,z]^T$, $T$ denotes a transpose, and $L({\bf x})$ represents the nonlinear functions and {\bf C} is a constant matrix. For the Lorenz system, 
\[F=\begin{pmatrix}
-\sigma & \sigma & 0 \\
r & -1 & 0 \\
0  & 0  & -b \\
\end{pmatrix},
L({\bf x})=\begin{pmatrix}
0 \\
-xy \\
xz  \\
\end{pmatrix},
{\bf C=0} \]
By an inspection of $F$ of the Lorenz system, we suggest that no  self-coupling is necessary, since all the diagonal elements are negative and hence all the error functions due to the self-coupling functions appear as negative definite in $\dot{V}(e)$ of Eq.~\eqref{Vex4}. This is supported by the condition of a critical self-coupling  $\epsilon_{1c}>-\sigma/2$ in \eqref{le4} that includes a condition, $\epsilon_1=\epsilon_{1c}=0$, as described above. However, we always apply one scalar self-coupling function to Lorenz systems so as to realize  first a locally stable CS state (MSF based stability). On the other hand,  a nonzero element exists in the upper triangle of $F$ that is connected to the $\dot x_{1,2}$ dynamics. This suggests an addition of one selective directed cross-coupling link defined by a coupling function involving the $y_{1,2}$ variable to the $x_2$  dynamics as shown in Eq.~\eqref{lor1} since the nonzero element is connected to the $y_{1,2}$ variable in $\dot x_{2}$ and it eliminates the additional error functions that appear in $\dot{V}(e)$ due to the presence of $y_{1,2}$ variable in the dynamical equations of the system. As example, $e_y$  is removed from Eq.~\eqref{ex3}  by the addition of a cross-coupling function (representing a coupling link) using the $y_{1,2}$-variable, and it suffices satisfying the condition, $\dot{V}(e)<0$. By a selection of two  linear self- and cross-coupling links between two Lorenz systems, the global stability of synchrony and other benefits such as the robustness of synchrony to large parameter perturbation, as described above, are  ensured. 
\par  This observation from Lorenz systems leads us to  define a  set of general conditions regarding the  selection of a coupling profile, between any two nodes of a network, from the LFM, $F$=\{$f_{ij}$\}, of any dynamical system that represents each node of the network. 
We always apply  a bidirectional self-coupling link first to realize a locally stable (MSF condition) synchronization. The question is which self-coupling  link  is most appropriate? Then what is the appropriate choice of the cross-coupling link? We frame a  general guideline that the diagonal elements of the LFM determine the choice of self-coupling links and the off-diagonal elements, in the upper triangle, determine  the choice of  cross-coupling links. Accordingly, we propose the coupling conditions, (1) {\it a primary choice of bidirectional self-coupling link is made for each positive diagonal element ($f_{ii}> 0$) in the LFM; for any negative diagonal element, self-coupling is redundant,} (2) {\it  if one or more zeros appear in the diagonal elements, at least one self-coupling link is  added to the related dynamical equation}, (3) {\it a  cross-coupling link is essential for each nonzero off-diagonal element ; however, no cross-coupling is needed if $f_{ij}$ =-$kf_{ji}$ where $k$ is a constant.} 
\par Above proposition clearly justifies our  selection of self-coupling and directed cross-coupling links in two Lorenz systems as demonstrated  above. We present more supporting examples with the paradigmatic HR neuron model \cite{m-2}, the SM laser model \cite{m-3},  the R\"ossler system \cite{m-4} and a Sprott system \cite{m-5} whose LFMs are ordered left to right, respectively,
\begin{equation}
F=\scriptscriptstyle{\begin{pmatrix}
	0 & 1 & -1\\
	0 & -1 & 0\\ 
	\mu r  & 0  & -\mu
	\end{pmatrix}};
\scriptscriptstyle{\begin{pmatrix}
 0 & 1 & 0\\
 r & -b & 0\\
 0 & 0 & -a 
	\end{pmatrix}};
\scriptscriptstyle{
	\begin{pmatrix}
0 & -1 & -1 \\
1 & a & 0 \\
0  & 0  & -c 
	\end{pmatrix}};
\scriptscriptstyle{
	\begin{pmatrix}
0 & -a & 0 \\
1 & 0 & 1 \\
r  & 0  & -1
\end{pmatrix}}. \label{LFM}
\end{equation}
\par For coupling two systems, we now proprose  coupling profiles for above four example models. For a HR system, the diagonal element $f_{11}$  of the LFM is 0 and hence we apply one bidirectional self-coupling link, involving the $x_{1,2}$-variables, to the dynamical equations of $x_1$; this is an appropriate  selection of the self-coupling link for two HR systems since all other diagonal elements ($f_{22}$ and $f_{33}$) are negative. In the upper triangle of the LFM, $f_{12}$ is 1 ($f_{12}\neq -f_{21}=0$) and hence one directed cross-coupling link involving the $y_{1,2}$-variables is to be added to the dynamics of $x_1$. No additional cross-coupling is needed since $f_{31}=-kf_{13}~ (k=\mu r)$. 
From the LFM of the SM laser system, second in the row, $f_{11}=0$, $f_{22}$ and $f_{33}$ are found negative, and the selection of self-coupling involves the $x_{1,2}$-dynamics and the cross-coupling ($f_{12}=1, f_{21}=r$)  involves $y_{1,2}$ variables and both are added to $x_1$-dynamics. No other cross-coupling is needed since other off-diagonal elements are $f_{13}=f_{23}=0$.   For the R\"ossler system  $f_{11}=0$ and $f_{22}=a$ as seen from the  LFM, third in the row, and hence we add one self-coupling involving the $y_{1,2}$-variables to the $y_{1,2}$-dynamics and one cross-coupling involving the $z_{1,2}$-variable to the dynamics of $x_1$- variable  ($f_{13}=-1; f_{31}=0$); no other cross-coupling is needed since $f_{12}$=-$f_{21}$. The LFM of the Sprott system is shown last in the  row. We have two options for bidirectional self-coupling ($f_{11}=f_{22}=0$, $f_{33}=-1$); one can select either of the pairs of state variables $x_{1,2}$ or $y_{1,2}$ witout any effect on the result. One cross-coupling involving the $z_{1,2}$ variable is added to the dynamics of $y_2$ since $f_{23}$=1 ($f_{32}=0$). No other cross-coupling is needed since $f_{12}=-af_{21}$. 
\par As a result, guided by the above conditions (1)-(3) and as elaborated above, we obtain the following self-coupling and cross-coupling matrices, $H_s$ and $H_c$, respectively, for building coupled systems using the  HR model, the SM model, the R\"ossler system and Sprott system, and presented in order from left to right,  respectively,
\begin{equation}
H_s=
{\begin{pmatrix}
	1 & 0 & 0\\
	0 & 0 & 0\\ 
	0   & 0  & 0
	\end{pmatrix}};
{\begin{pmatrix}
	1 & 0 & 0\\
	0 & 0 & 0\\ 
	0   & 0  & 0
	\end{pmatrix}};
{ \begin{pmatrix}
	0 & 0 & 0 \\
	0 & 1 & 0 \\
	0  & 0  & 0 
	\end{pmatrix}};
{	\begin{pmatrix}
	1 & 0 & 0\\
	0 & 0 & 0\\ 
	0   & 0  & 0
	\end{pmatrix}} \label{Hs}
\end{equation}
and
\begin{equation}
H_c=
{\begin{pmatrix}
	0 & 1 & 0\\
	0 & 0 & 0\\ 
	0   & 0  & 0
	\end{pmatrix}};
{\begin{pmatrix}
	0 & 1 & 0\\
	0 & 0 & 0\\ 
	0   & 0  & 0
	\end{pmatrix}}; 
{  \begin{pmatrix}
	0 & 0 & 1 \\
	0 & 0 & 0 \\
	0  & 0  & 0 
	\end{pmatrix}};
{
	\begin{pmatrix}
	0 & 0 & 0\\
	0 & 0 & 1\\ 
	0   & 0  & 0
	\end{pmatrix}} \label{Hc}
\end{equation}
 The elements of the coupling matrices $H_s$ and $H_c$,
in Eqs.~\eqref{Hs} and \eqref{Hc}, respectively, are chosen accordingly, \\
(1) {\it HR system: one self-coupling link and one cross-coupling link},  all elements of the coupling matrices are 0 except ${H_s}_{11}$=1 and ${H_c}_{12}$=1,\\
 (2) {\it SM system: one self-coupling link and one cross-coupling link}, all elements of the coupling matrices are 0 except ${H_s}_{11}$=1 and ${H_c}_{12}$=1,\\
 (3) {\it R\"ossler system: one self-coupling link  and one cross-coupling link}, all elements of the coupling matrices  are 0 except ${H_s}_{22}$=1 and ${H_c}_{13}$=1,\\
 (4) {\it Sprott system: one self-coupling link and one cross-coupling link.} All elements of the coupling matrices are 0 except ${H_s}_{11}$=1 and ${H_c}_{23}$=1.
  \par
\begin{figure}[!ht]
  	\centering
  	\hspace{0pt}
  	\includegraphic[width=2.5cm,height=3.25cm] {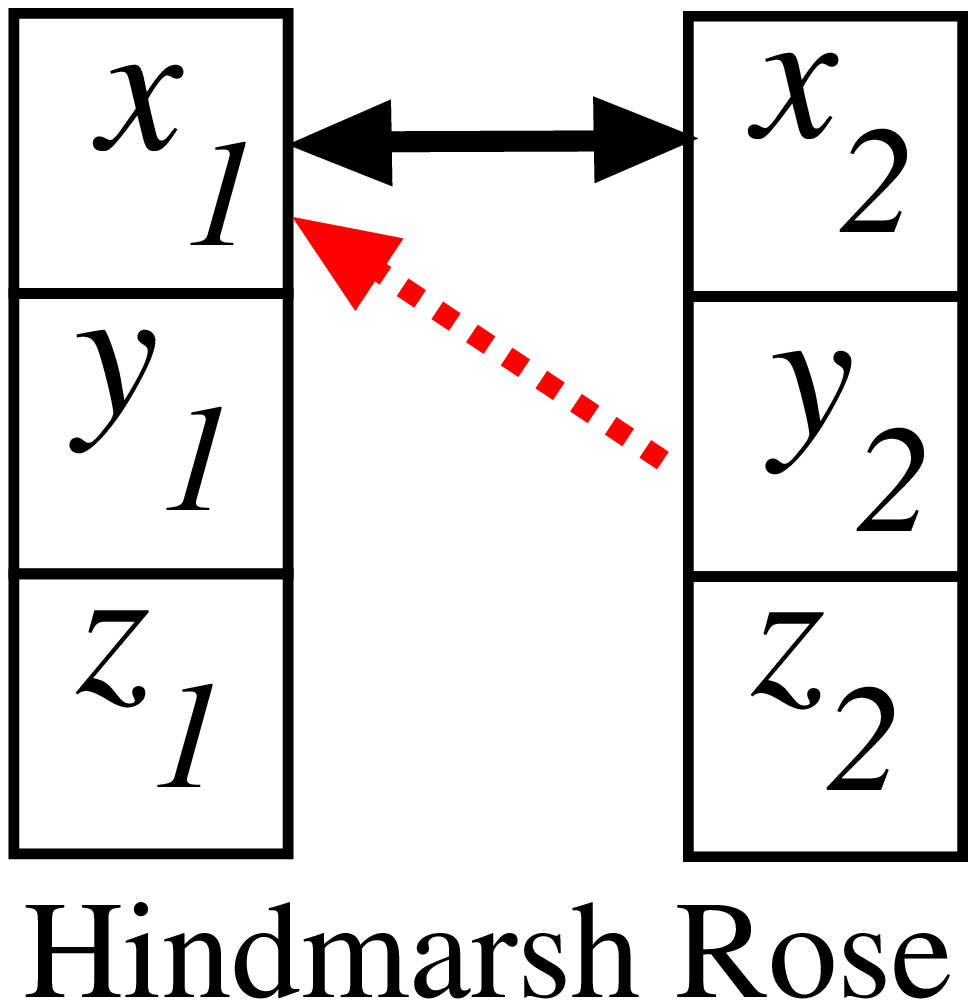}{(a)}{0pt}{0} 
  	\hspace{-5pt}
  	\includegraphic[width=3cm,height=3cm]{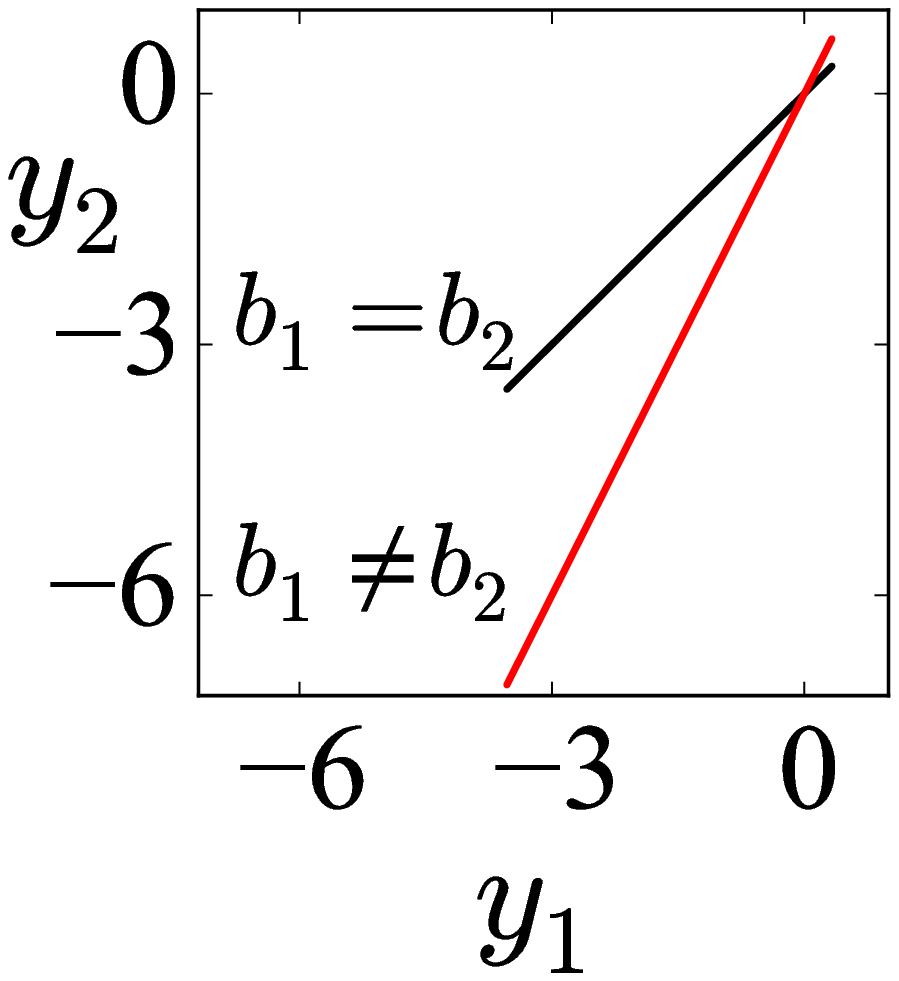}{(e)}{0pt}{0}\\
  	\includegraphic[width=2.5cm,height=3.25cm]{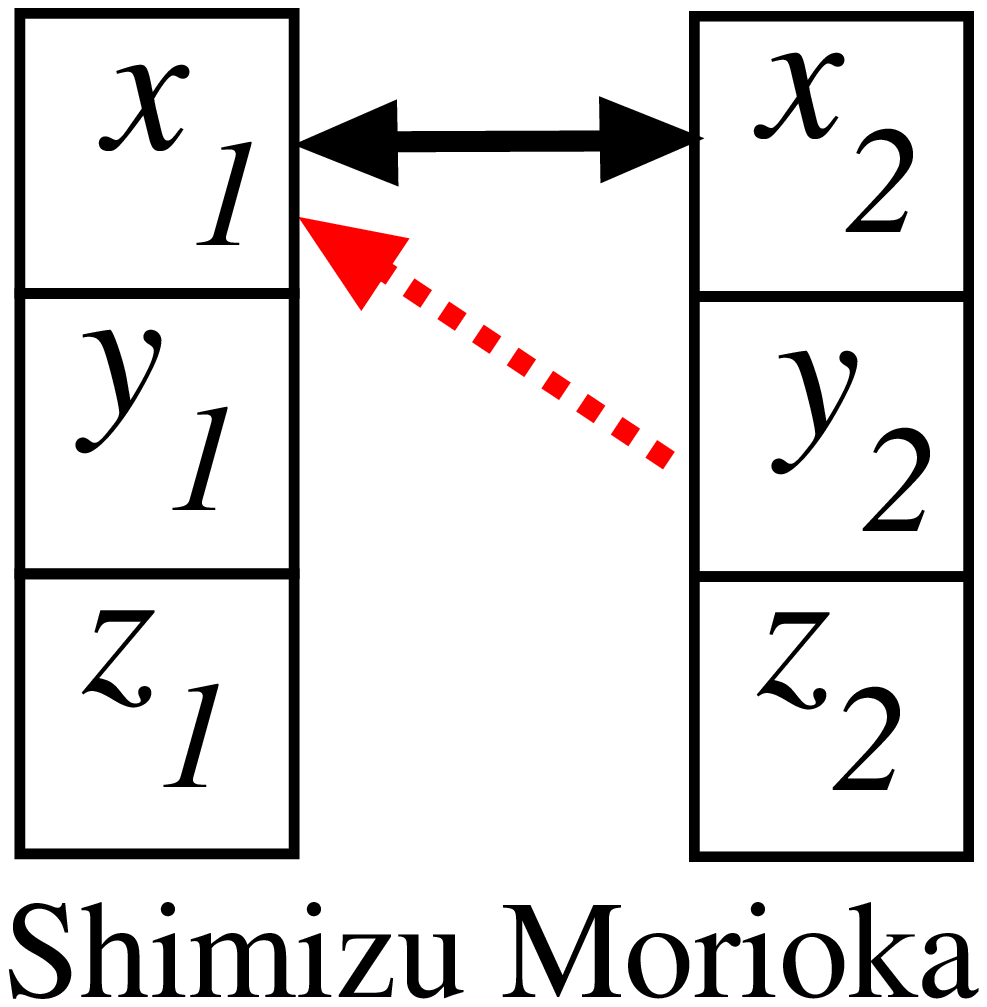}{(b)}{0pt}{0}
  	\hspace{-5pt}
  	\includegraphic[width=3cm,height=3cm]{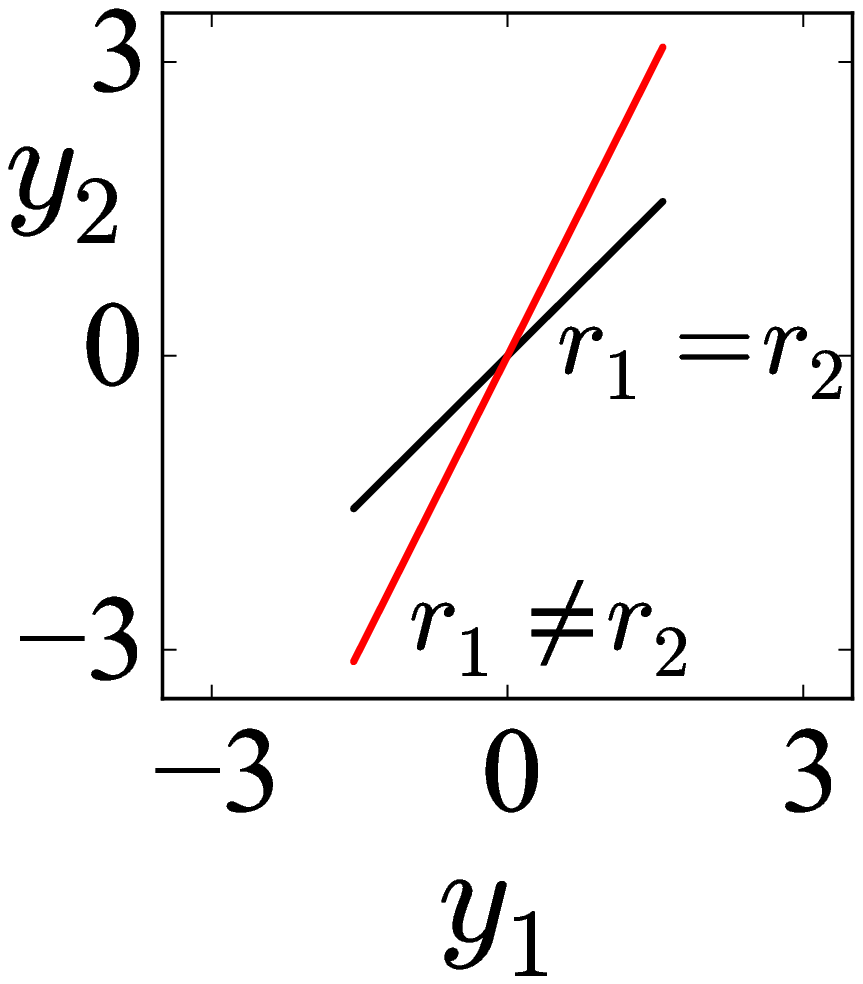}{(f)}{0pt}{0} \\
  	\hspace{0pt}
  	\includegraphic[width=2.5cm,height=3.25cm]{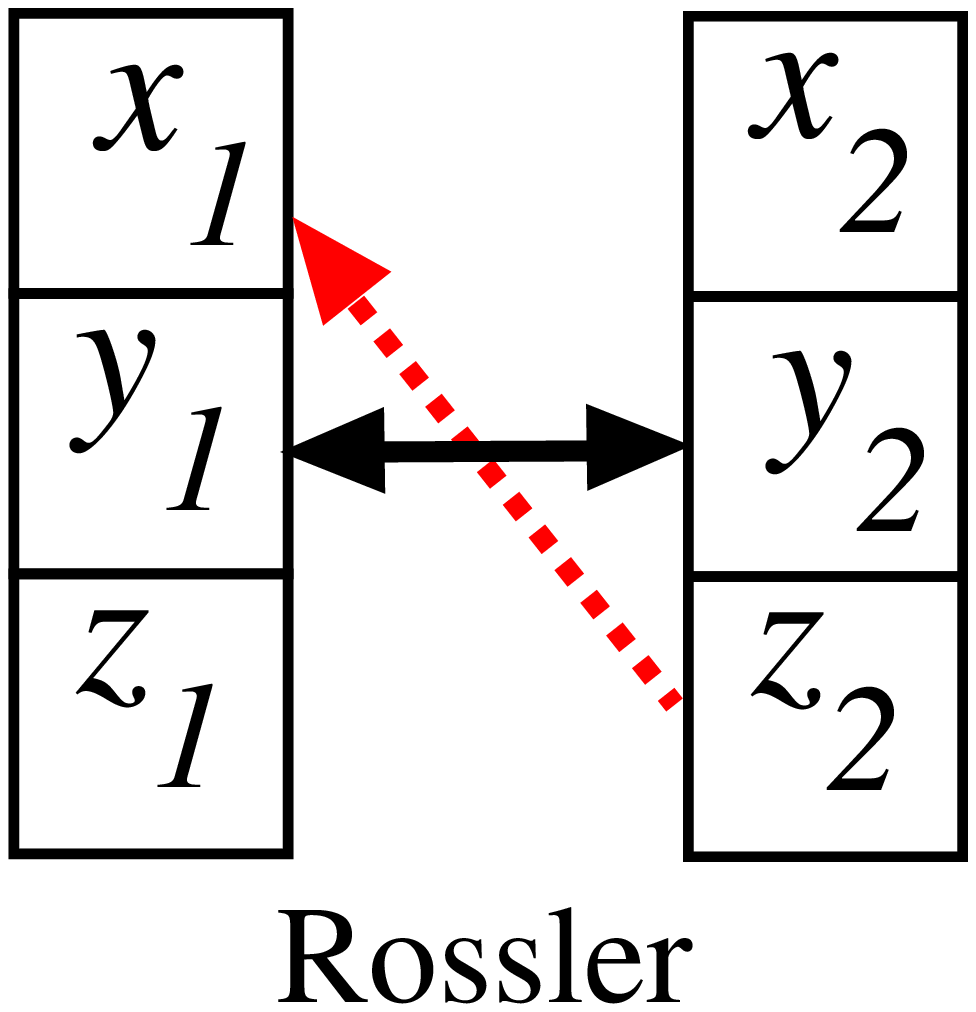}{(c)}{0pt}{0}
  	\hspace{-5pt}
  	\includegraphic[width=3cm,height=3cm]{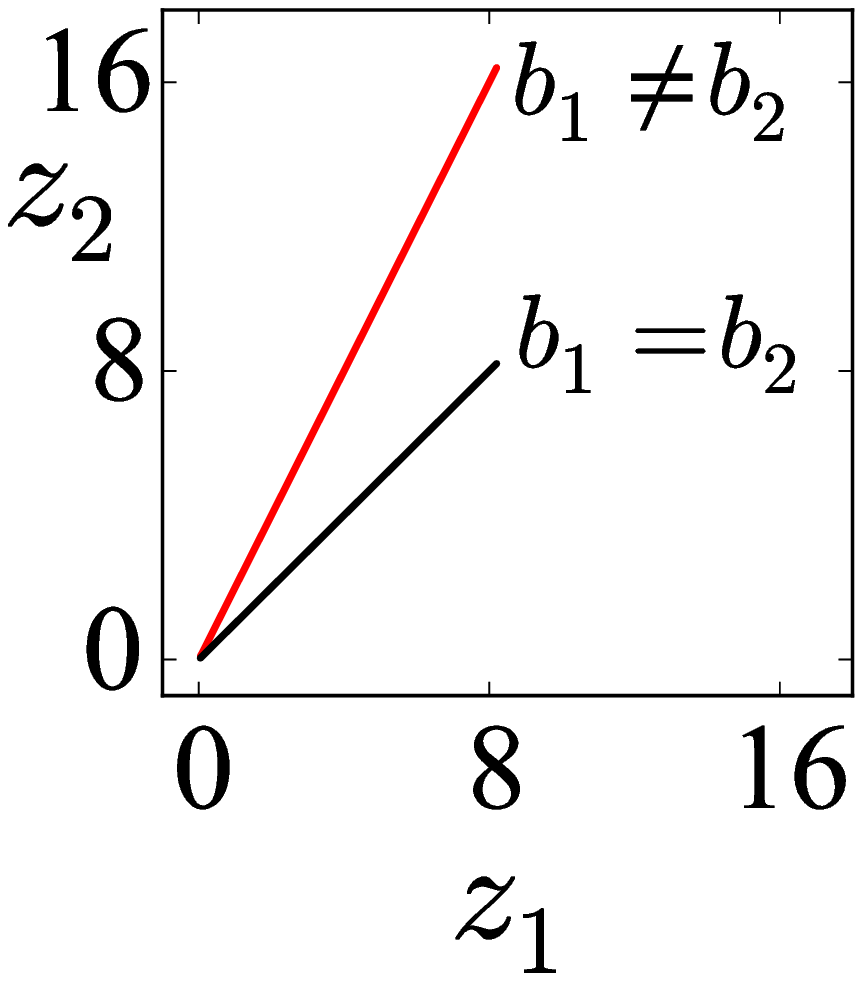}{(g)}{0pt}{0}\\
  	\hspace{5pt}
  	\includegraphic[width=2.5cm,height=3.25cm]{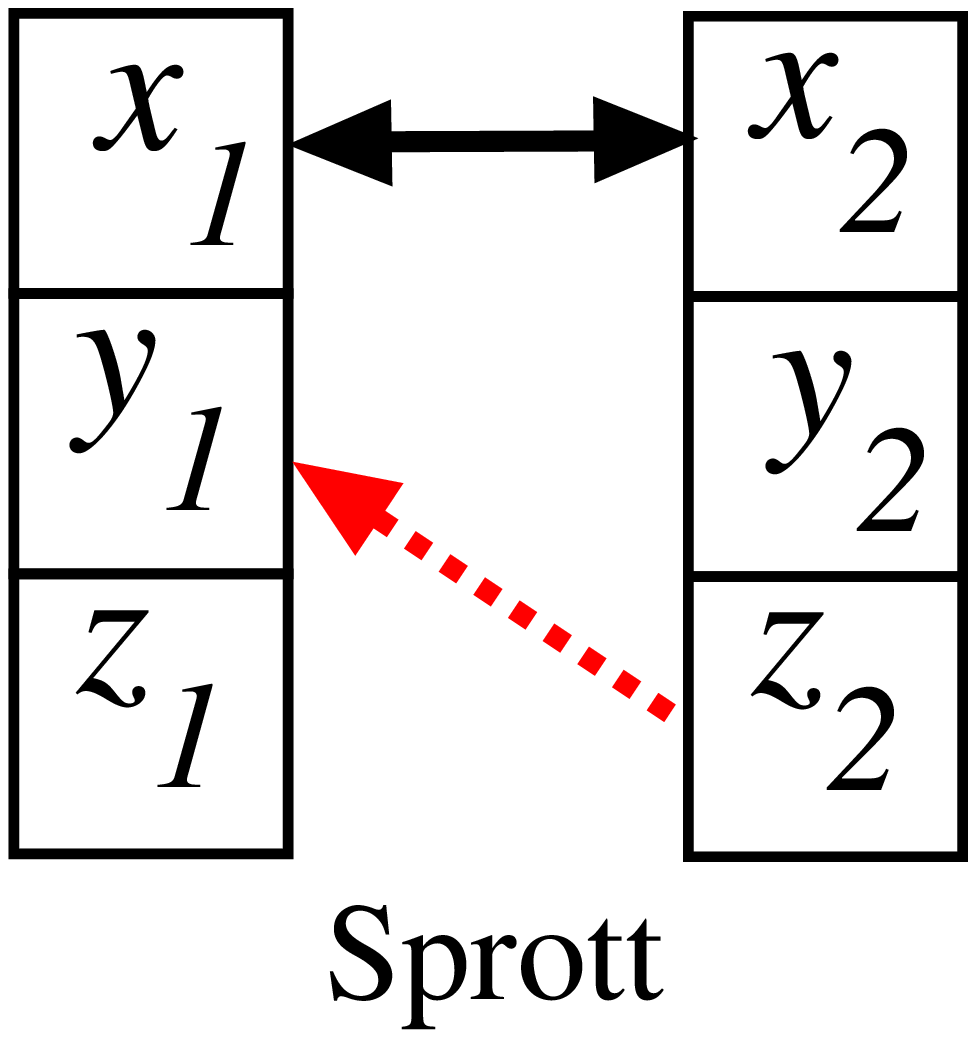}{(d)}{0pt}{0}
  	\hspace{-5pt}
  	\includegraphic[width=3cm,height=3cm]{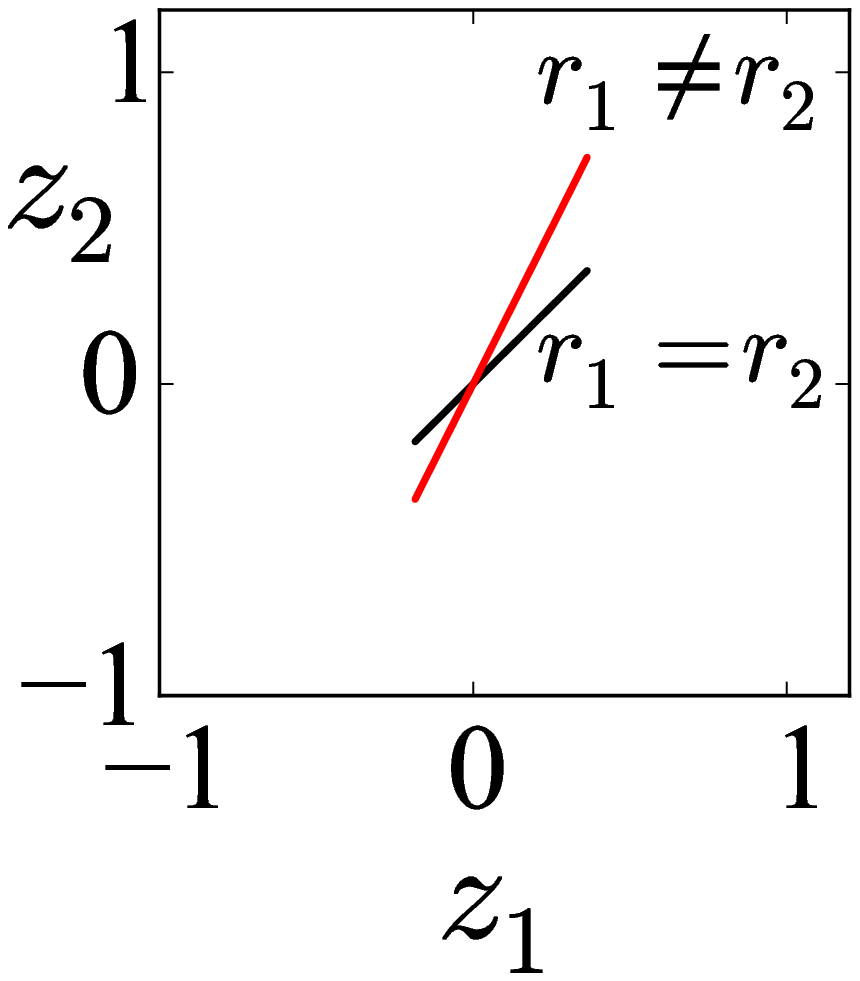}{(h)}{0pt}{0}
  	\vspace{-5pt}
  	\caption{(Color online). Coupling profiles for coupled HR and SM models, R\"ossler and Sprott systems in (a)-(d), respectively. $X_i$=$[x_i, y_i, z_i]^T$ state variables of $i$th oscillator. Solid arrows (black) for bidirectional self-coupling links, dashed  arrows (red) for directed cross-coupling links. Coupling strengths for HR ($b_1=0.5$, $b_2=1$): $\epsilon_1=0.45$, $\epsilon_2=1$; SM ($r_1$=0.5, $r_2$=1): $\epsilon_1=0.1$, $\epsilon_2=1$; R\"ossler system ($b_1$=0.1, $b_2$=0.2): $\epsilon_1=0.21$, $\epsilon_2=-1$; Sprott system ($r_1$=0.5, $r_2$=1): $\epsilon_1=0.1$, $\epsilon_2=1$.  Synchronization manifolds for coupled (e) HR model, (f) SM model, (g) R\"ossler system, (h) Sprott system: identical systems in black, detuned systems in gray (red).  Other parameters are given in \cite{m-2,m-3,m-4,m-5}.} 
  	 	\label{fig2}
  \end{figure}

 The coupling profiles are pictorially demonstrated in Fig.~\ref{fig2}(a)-(d) for the above  systems that always consist of one self-coupling (black arrow) and one cross-coupling link (red arrow) for 2-node coupled systems. The connectivity matrices (self- and cross-coupling links) are only different for different systems.
  
 \section{Two coupled systems} 
\par 
\par  The proposed selection of  coupling profile  favors global stability of synchrony in two-coupled  systems. Numerical results are summarized for all the systems in the immediate right panels of Figs.~\ref{fig2}(e)-(h). 
 Analytical details are provided for the  HR model in this section  and for rest of the models see the Appendix. 
  For each coupled system, the coupling profile  is selected, as suggested above, to  realize a globally stable CS for identical systems. CS manifolds of identical oscillators are shown in black lines and the GS manifolds of the detuned systems are shown in red lines. CS to GS transition due to a parameter perturbation is reflected by a  rotation of the synchronization manifold for a similar reason as explained above for the Lorenz system. The amplified response of the detuned systems' attractor \cite{Suman} under a perturbation  is explained further using the example of the slow-fast HR systems. 
\par 
The coupling profile ($H_s$  and  $H_c$) is taken from  the Eqs.~\eqref{Hs}-\eqref{Hc} for the HR neuron model that  suggests one bidirectional self-coupling and one directed cross-coupling for globally stable synchrony. 
Accordingly, two coupled HR systems  appears as,
	\begin{equation}
	\begin{array}{l}
	\dot{x}_{1}=a x_{1}^2 -x_{1}^3+{y}_{1} -z_{1}+I+ \epsilon_1(x_{2}-x_{1})+\epsilon_2(y_{2}-y_{1})\\
	\dot{x}_{2}=a x_{2}^2 -x_{2}^3+{y}_{2} -z_{2}+I+ \epsilon_1(x_{1}-x_{2})\\
	\dot{y}_{1,2} =b_{1,2}(1-dx_{1,2}^2)-y_{1,2}  \\ 
	\dot{z}_{1,2}=\mu [r (x_{1,2}-c)-z_{1,2}]
	\end{array} \label{s4}
	\end{equation}
The bidirectional self-coupling link $(x_{2,1}-x_{1,2})$ is added to the dynamics of $x_{1,2}$-variables while the  cross-coupling link $(y_2-y_1)$ is directed and added only to the dynamics of  $x_1$ of the first oscillator. This choice can be reversed, i.e., ($y_1-y_2)$ could be added alternatively to the dynamics of $x_2$ without any effect on the final results. 
\par  The evolution of the error functions   $\bf e$=$[e_x,e_y,e_z]^T=[x_1-x_2,y_1-y_2,z_1-z_2]^T$ is given by,
	\begin{equation}
	\begin{array}{l}
	\dot{e}_x=ae_xe_p-\frac{e_x^3}{4}-\frac{3}{4}e_xe_p^2+e_y-e_z-2\epsilon_1e_x-\epsilon_2e_y,\\
	\dot{e}_y=b_1(1-dx_1^2)-b_2(1-dx_2^2) -e_y, \\
	\dot{e}_z=\mu r e_x - \mu e_z \label{s5},
	\end{array}
	\end{equation}
	where, $e_p=x_1+x_2$ so that $x_1^2-x_2^2=e_xe_p$ and $x_1^3-x_2^3=\frac{e_x}{4}(e_x^2+3e_p^2)$.
	For a global stability of  $(e_x=0,e_y=0,e_z=0)$, we consider a Lyapunov function, $V(e)=\frac{1}{2}e_x^2+\frac{1}{2}e_y^2+\frac{1}{2\mu r}e_z^2$.
	We first check the stability of $x_1=x_2$ and $z_1=z_2$, separately, by defining a Lyapunov function, $V'(e_x,e_z)=\frac{1}{2}e_x^2+\frac{1}{2\mu r}e_z^2$ when its time derivative is
	\begin{align}
	\dot{V'}(e_x,e_z)=&-e_x^2(-ae_p+2\epsilon_1+\frac{3}{4}e_p^2)\nonumber \\ 
	&-\frac{e_x^4}{4}+(1-\epsilon_2)e_ye_x - \frac{e_z^2}{r}.\label{s6}
	\end{align}
	$\dot V'$($e_x, e_z)<0$ provided  $\epsilon_2=1$ and $\frac{3}{4}e_p^2-ae_p+2\epsilon_1\geq 0$. 
To satisfy the condition, we derive the roots of the equation,
\begin{equation}
\frac{3}{4}e_p^2-ae_p+2\epsilon_1=0 \label{s7}
\end{equation}  
which are given by
\begin{equation}
{e_p}_{1,2} =\frac{2a\pm\sqrt{a^2-6\epsilon_1}}{3}. \label{s8}
\end{equation}
${e_p}_{1,2}$ will now be positive real if $a^2-6\epsilon_1 \geq0$ that implies ${\epsilon_1}\leq\frac{a^2}{6}$. Thus 
  $\dot{V'}(e_x,e_z)<0$ is satisfied when ${\epsilon_{2}}=1$ and ${\epsilon_1}\leq a^2/6$ and 
this implies $x_1=x_2$ and $z_1=z_2$ is asymptotically stable as $t\rightarrow\infty$. Substituting the  condition in Eq.~\eqref{s5} and assuming identical  systems ($b_1=b_2$), the error dynamics  $\dot{e}_y=-e_y$ is found when,
\begin{equation}
	\dot{V}(e_x,e_y,e_z)=-\frac{e_x^4}{4} - e_y^2 - \frac{e_z^2}{r} <0 \label{s9}
\end{equation}
for ${\epsilon_1}\leq \frac{a^2}{6}$ and ${\epsilon_{2}}=1$. 
The synchronous state $x_1=x_2, y_1=y_2$ and $z_1=z_2$, basically a CS state of the coupled identical HR system, is now globally stable. 
\par
\begin{figure}[!ht]
    \centering
	\includegraphic[width=3.5cm,height=3.5cm]{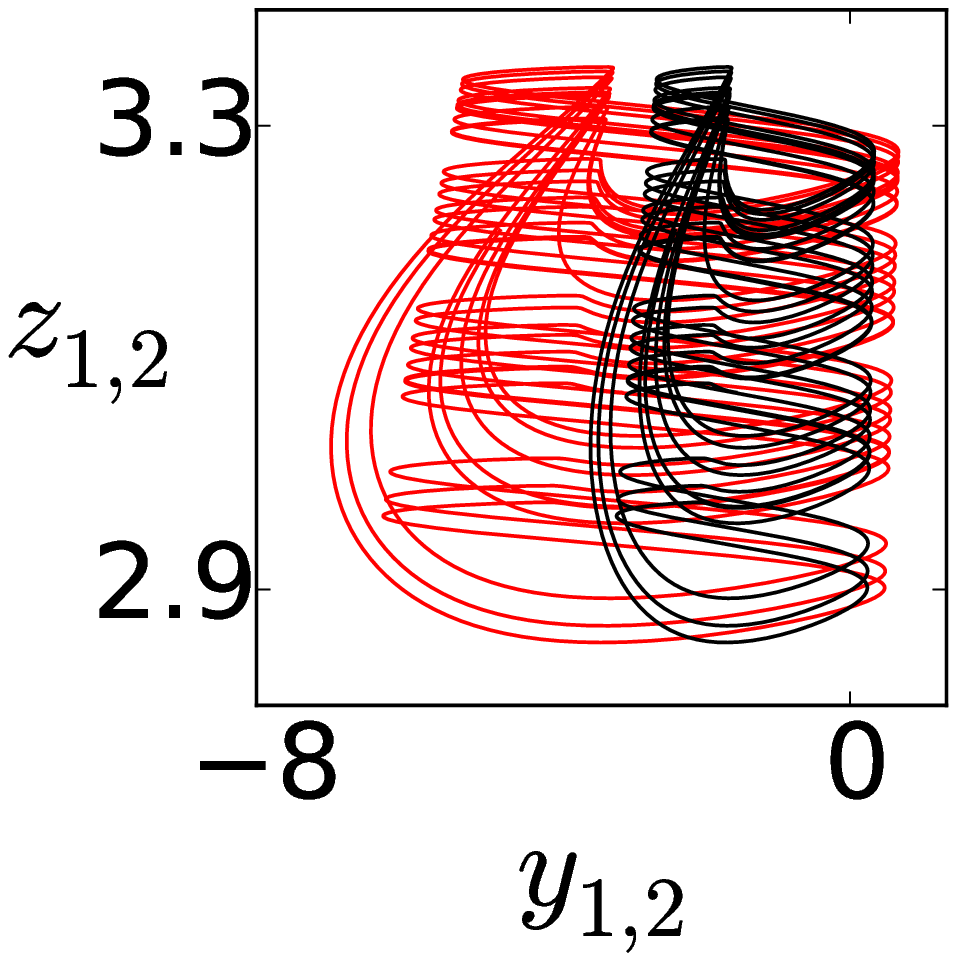}{(a)}{0pt}{0}
	\includegraphic[width=3.5cm,height=3.5cm]{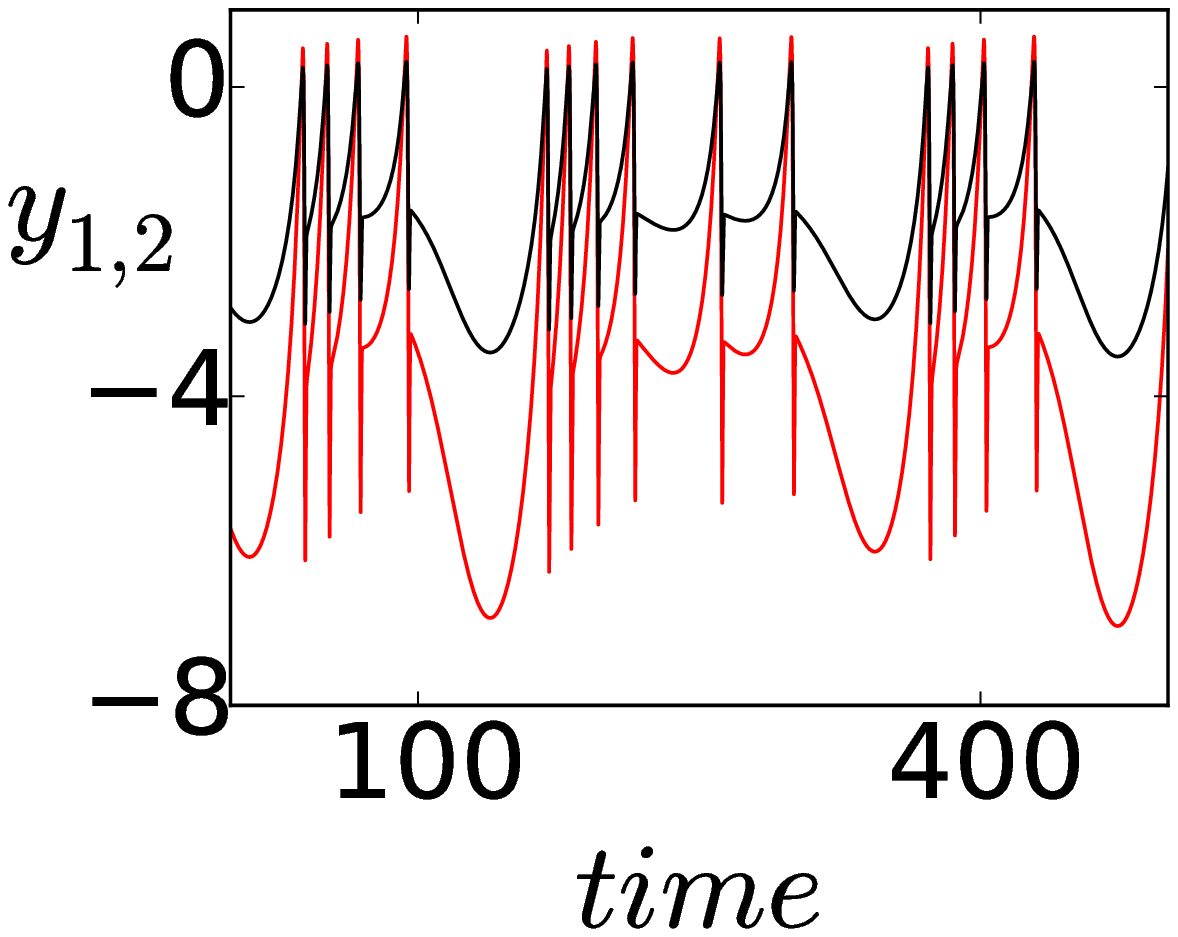}{(b)}{0pt}{0}
	\includegraphic[width=3.5cm,height=3.5cm]{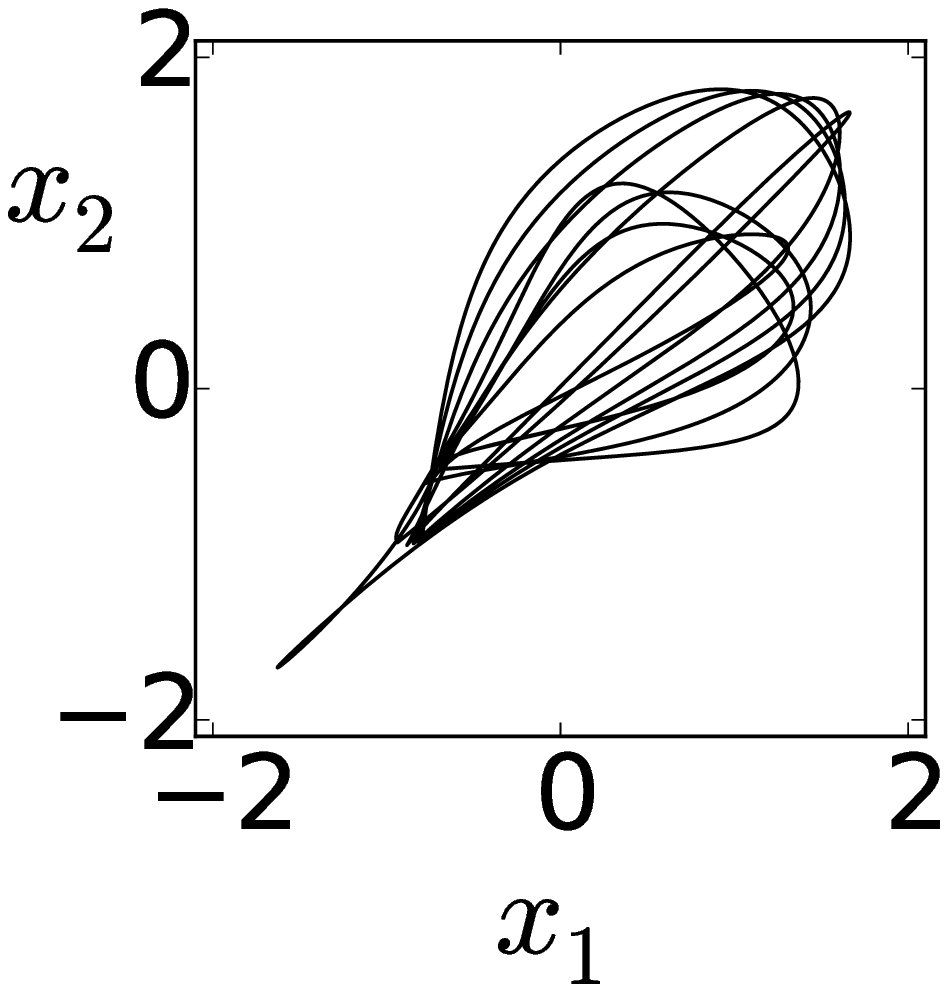}{(c)}{0pt}{0}
	\hspace{-0pt}
	\includegraphic[width=3.5cm,height=3.5cm]{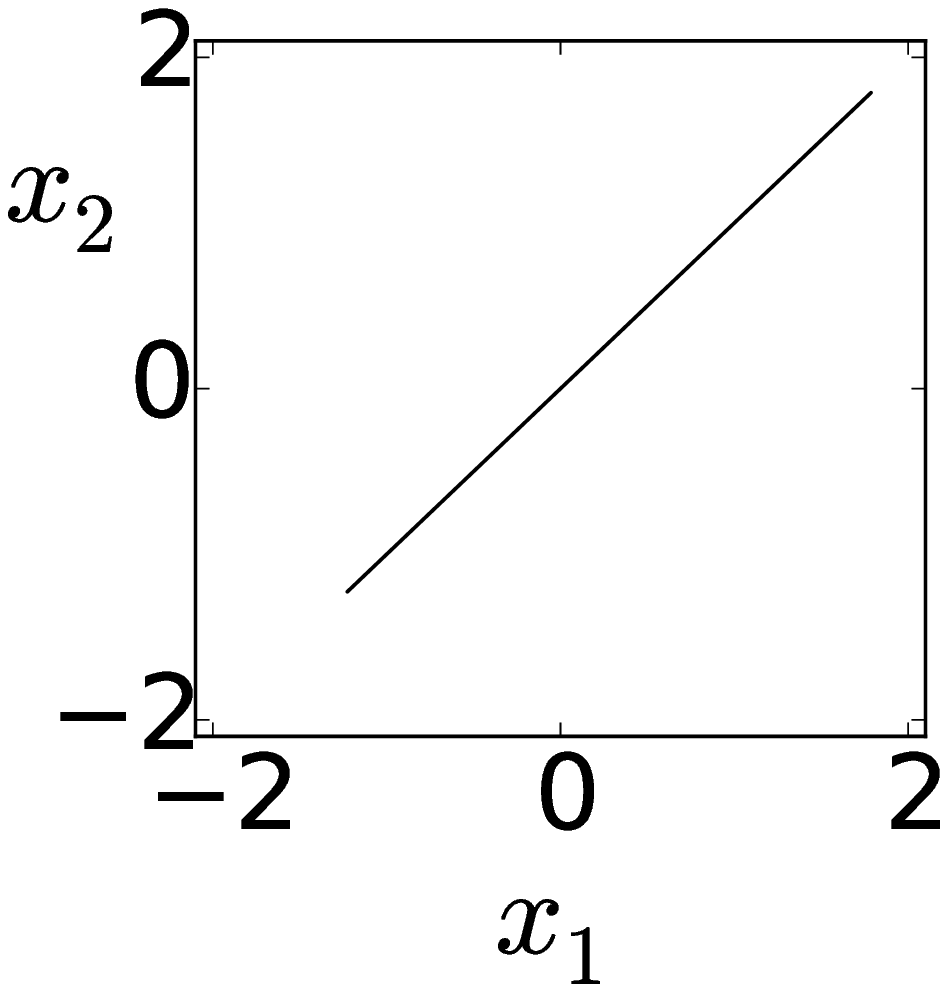}{(d)}{0pt}{0}
	\hspace{0pt}
\caption{(Color online) Globally stable synchrony by  addition of cross-coupling link in two coupled HR systems. Plot of attractors $y_{1,2}$ vs $z_{1,2}$ (a) confirms expansion of the detuned attractor (red) along the $y$-direction compared to unperturbed attractor (black), (b) time series of $y_{1,2}$ showing amplification. Other variables remain in CS and not shown here. For detuning  $b_1$=0.5, $b_2$=1 are taken. Plots of $x_2$ vs. $x_1$, (c)  identical  systems ($b_2=b_1$=1),	with a drifting of coupling strength ($\epsilon_1=0.45<e_{MSF}$),   
 (d) identical systems after adding a cross-coupling  link ($\epsilon_2$=1). }
	\vspace{-5pt}\label{sf1}
\end{figure}

 Now the effect of heterogeneity on the stability of CS is tested by detuning the parameter, ($b_1\neq b_2$), other parameters are kept unchanged.
The induced heterogeneity has no effect on Eq.\eqref{s6} and related conditions and hence the stability  of  $x_1=x_2$ and $z_1=z_2$ is still preserved. The stability of $e_y$ is only to be checked against a parameter detuning. The equation of $\dot{e}_y$ after a perturbation is written from Eq.\eqref{s5},
	\begin{equation}
	\begin{array}{l}
	\dot{e}_y =b_1(1-dx_1^2)-b_2(1-dx_2^2)-e_y \\ \quad\; = (b_1-b_2)(1-dx_1^2)-e_y 
	\; =\frac{b_1-b_2}{b_1}(\dot{y}_1+y_1)-e_y, 
	\end{array} \label{s9}
	\end{equation}
and this leads to
	\begin{equation}
	\begin{array}{l}
	\dot{y}_1(1-\frac{b_1-b_2}{b_1})-\dot{y}_2
	=-y_1(1-\frac{b_1-b_2}{b_1})+y_2  \\
	\dot{y}_1\frac{b_2}{b_1}-\dot{y}_2=-(y_1\frac{b_2}{b_1}-y_2).\\
	\end{array}\label{s10}
	\end{equation}
From \eqref{s10}, the revised error dynamics is 
$\dot{e}_y^*=-e_y^*$, 
 where  $e_y^*=y_1\frac{b_2}{b_1}-y_2$ is the modified error function. 
Accordingly, the Lyapunov function is redefined in terms of the modified error functions whose time derivative is
\begin{equation}
	\dot{V}^{*}(e_x,e_y^*,e_z)=-\frac{e_x^4}{4}-e_y^{*2}- \frac{e_z^2}{r} <0 \label{s12}
\end{equation}  
and valid for same conditions, ${\epsilon_1}\leq\frac{a^2}{6}$ and  ${\epsilon_2}=1$ as given above. It confirms an emergence of  a new globally stable synchronous state, $x_1=x_2$, $y_1 =\frac{b_1} {b_2}y_2$, $z_1=z_2$, due to parameter detuning, which we call as a GS state.  This is manifested by a rotation of the CS hyperplane ($y_1=y_2$, black line) to a GS hyperplane ($y_1=\frac{b_1}{b_2}y_2$, red line) as shown in Fig.~\ref{fig2}(e) where the angle of rotation is decided by the amount of detuning. For any choice of $b_1$ and $b_2$, the attractor of the detuned system shall be amplified or attenuated along the $y$-direction as determined by the amount of heterogeneity, i.e., by the ratio of (${b_1}/{b_2}$) as  shown in Fig.~\ref{sf1}(a).
The $y_2$-variable (red line) is only amplified twice ($y_2=2y_1$; $y_1$ in black line) in  Fig.~\ref{sf1}(b) for a choice of $b_1=0.5$ and $b_2=1.0$. A complete in-phase synchrony is maintained between $y_1$  and $y_2$.
 
Another important point is that synchrony is lost under a purely self-coupling ($\epsilon_2=0$) if its strength drifts below $\epsilon_1<e_{MSF}$, however, restored by the addition of one directed cross-coupling link as shown in $x_1$ vs. $x_2$ plots in Figs.~\ref{sf1}(c)-\ref{sf1}(d) since the range of $\epsilon_1$ ($\leq\frac{a^2}{6}$) is expanded.  
\par 

\section{Network Motifs}
 \par We focus here on network motifs  which are building blocks of many real world networks \cite{Alon} and check if the proposed general conditions for the selection of coupling profiles work in favor of global stability of synchrony. A few examples are presented that support our proposition and many more are cited in the Supplementary material \cite{SM}. The dynamics of the $i$-th oscillator in a network of N-oscillators under  self-coupling as well as cross-coupling is, 
 \begin{equation}
 \dot{\textbf{x}}_i= g(\textbf{x}_i) +  \epsilon_1\sum_{j=1}^N A_{ij} H_s(\textbf{x}_j-\textbf{x}_i)+\epsilon_2 \sum_{j=1}^N B_{ij}H_c(\textbf{x}_j-\textbf{x}_i) \label{s3},
 \end{equation}
 $g({\bf x_i})$ denotes the flow of the $i^{th}$ node in isolation, $i\in \{1, ..., N\}$, ${\bf x_i}\in \mathbb{R}^n$ is the state vector.   $\epsilon_1$ and $\epsilon_2$ are the  strength of  self- and cross-coupling links, respectively, between any two nodes. $A=\{A_{ij}\} \in \mathbb{R}^{N\times N}$ is the adjacency matrix  that defines the network topology via self-coupling links and $B=\{B_{ij}\} \in \mathbb{R}^{N\times N}$ is the connectivity matrix of the cross-coupling links of the network; $A_{ij}$=1 and $B_{ij}$=1, if $i^{th}$ node is connected to the $j^{th}$ ($j\neq i$) and 0 otherwise. $H_s$ and $H_c$ a usual define the self- and cross-coupling matrices  for any pair of nodes. 
\par We make appropriate selection of   $H_s$ and $H_c$ between any two nodes of a network as guided by the LFM of dynamical systems representing each node. 
We are allowed with $N-1$ directed cross-coupling links from any arbitrarily chosen node to all other nodes that defines the connectivity matrix $B$ of cross-coupling links. A restriction is imposed that the addition of cross-coupling links must not make an original directed link  undirected. In other words, $B$ must not change the original topology of a network-motif defined by $A$. 
The single node from where the outgoing directed cross-coupling links  are connected to all other nodes, maintains the isolated dynamics at a globally stable CS state when all coupling functions vanish for identical nodes. 
This arbitrarily chosen node effectively plays the role of a driving oscillator and, obviously, maintains the isolated dynamics. In the case of a parameter perturbation, the network emerges into a new coherent state  of  GS type, as explained above,  however, continues with the original dynamics so long as the newly emerged driver node's parameter is not perturbed as discussed above for two Lorenz systems.  
\par Using a few examples of network-motifs, the general  applicability of our coupling profile scheme  is establised analytically as well as numerically. All the network-motifs are drawn with self-coupling links in black arrows and cross-coupling links in  dashed arrows (red). Results are summarized here;  see the Supplementary material \cite{SM} for analytical details and more examples.
\subsection{Network motif: Shimizu-Morioka laser model}\label{sec:NM}
\begin{figure}[!ht]
	\centering
	\begin{minipage}{.2\textwidth}
		\hspace{-55pt}
		\includegraphic[width=2.75cm,height=2.75cm]{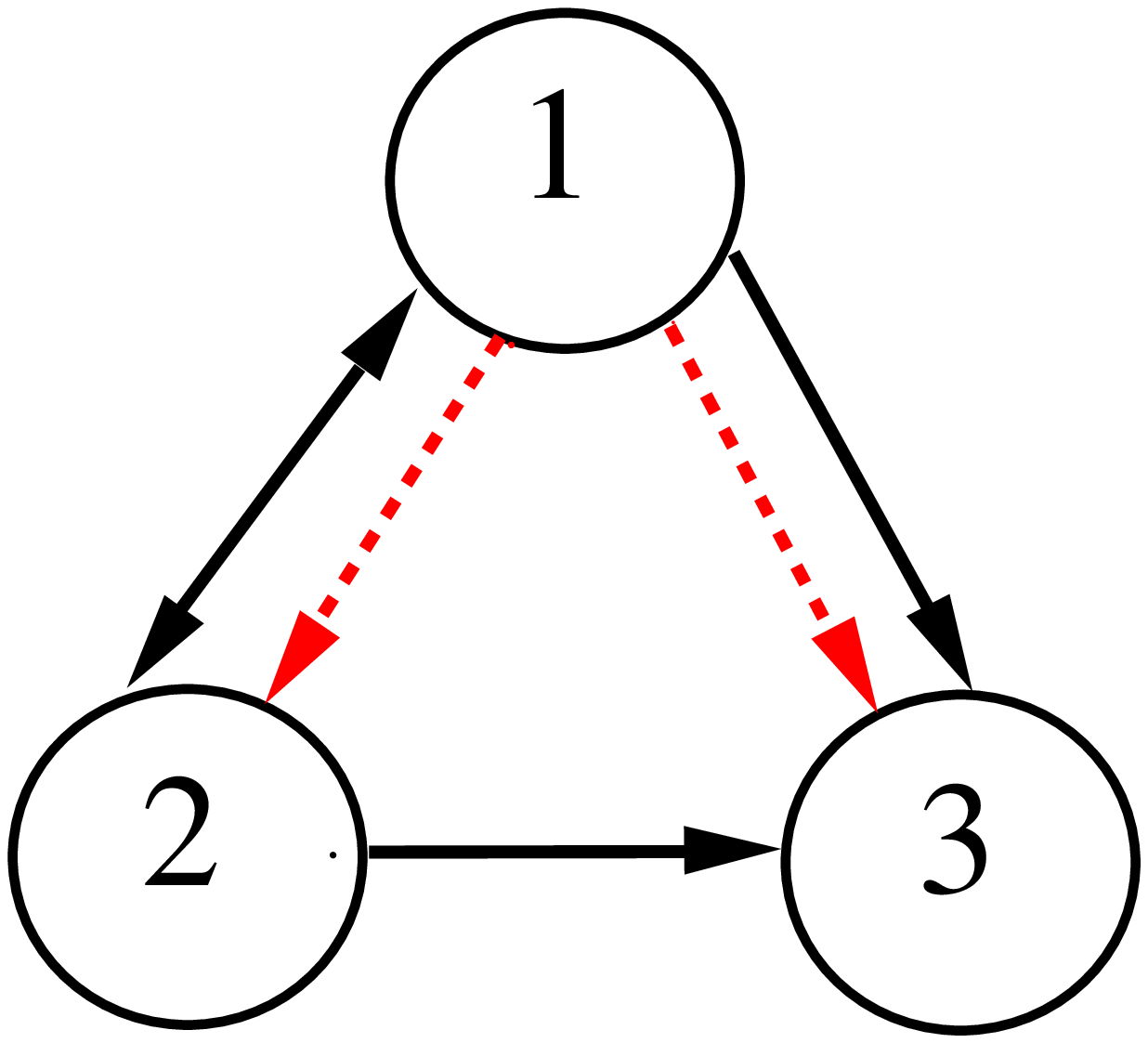}{(a)}{0pt}{0}
	\end{minipage}
	\begin{minipage}{.2\textwidth}
		\hspace{-45pt}
		\includegraphic[width=4.75cm,height=4cm]{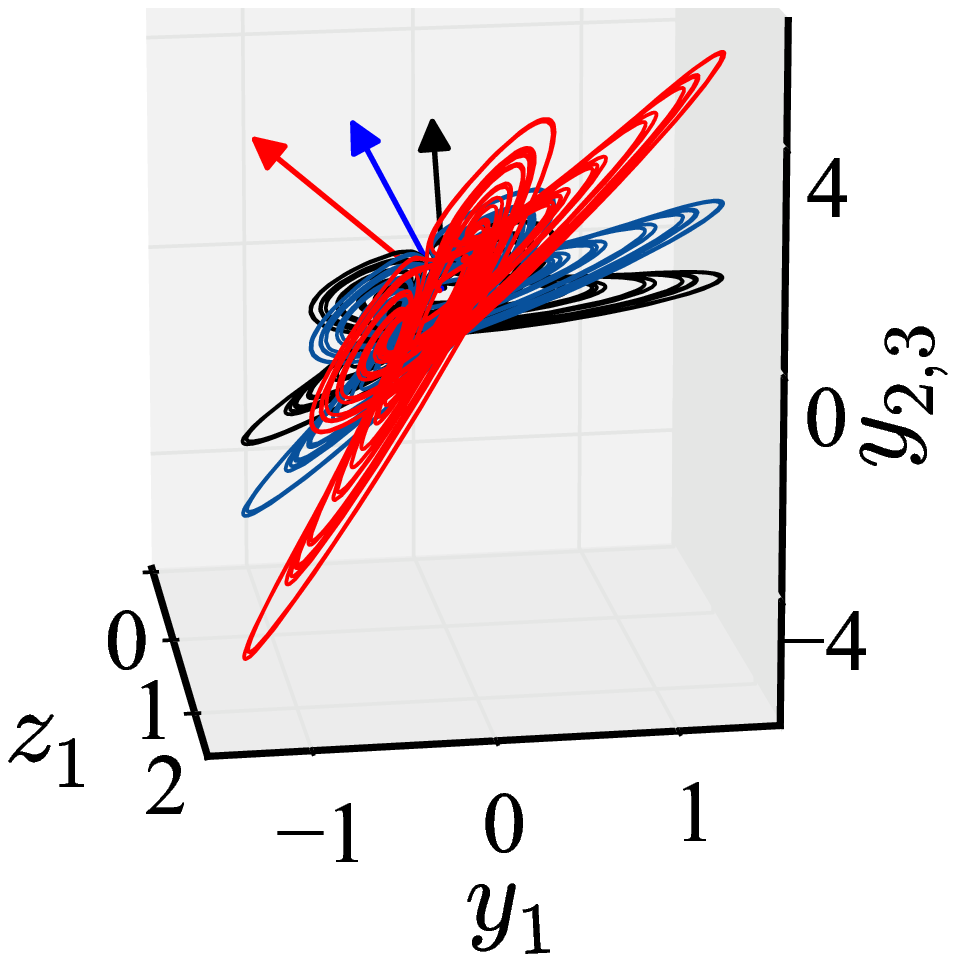}{(b)}{0pt}{0}
	\end{minipage}
	\caption{(Color online) (a) 3-node network motif of SM model. (b) Projection of synchronization hyperplanes. Parameters are $a$=0.375, $b$=0.826, $r_1$=1, $r_2$=2, $r_3$=4, $\epsilon_1$=0.1 and $\epsilon_2$=1.}
	\vspace{-5pt}\label{fig4ab}
\end{figure}

 A particular 3-node network motif is first shown in Fig.~\ref{fig4ab}(a) whose each node is represented by the SM laser model. The network topology $A$ is denoted by the self-coupling links in black arrows  and the connectivity matrix of cross-coupling links in dashed arrows (red) is represented by $B$,   
\begin{equation}
A=\left(
\begin{array}{rrr}
0 & 1 & 0 \\
1 & 0 & 0 \\
1 & 1 & 0 \\
\end{array} \right), 
B=\left(
\begin{array}{rrr}
0 & 0 & 0 \\
1 & 0 & 0 \\
1 & 0 & 0 \\
\end{array} \right).\label{s15}
\end{equation}
\begin{figure}[!ht]
	\centering
	\hspace{-10pt}
	\includegraphic[width=4cm,height=4cm]{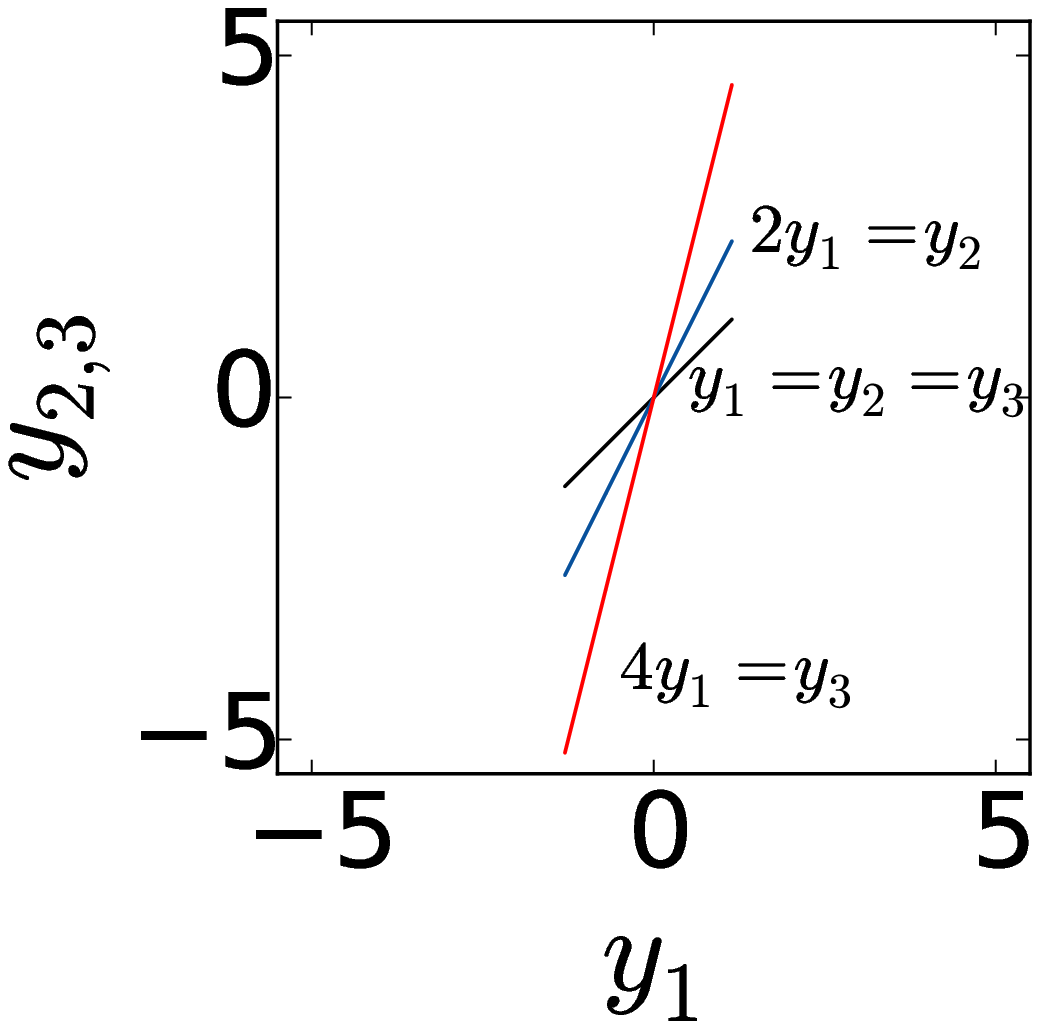}{(a)}{0pt}{0}
	\hspace{-5pt}
	\includegraphic[width=4cm,height=4cm]{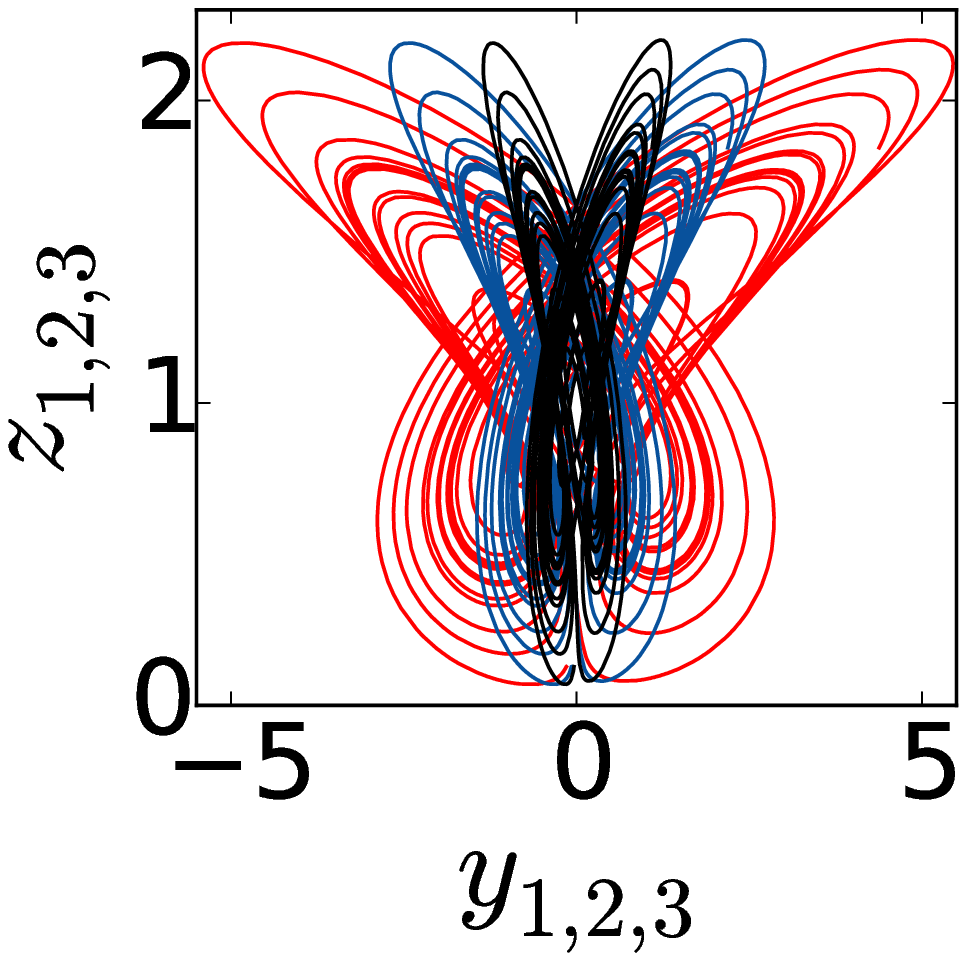}{(b)}{0pt}{0}
	\caption{(Color online) 3-node network-motif of SM model :  Synchronous manifolds  in, (a) all oscillators collapse to the reference manifold (black line) for identical parameters; for detuning two oscillators, their attractors rotated away, in gray (red line) and dark gray (blue line). Three attractors are plotted in, (b) unperturbed attractors in black, others in colors (blue, red). }	\vspace{-5pt}\label{sf3}
\end{figure}
Notice that the cross-coupling links do not change the network topology. $H_s$ and $H_c$ are obtained from Eqs.\eqref{Hs}-\eqref{Hc}, for the SM model representing each node,
\begin{equation}
{H_s}=\begin{pmatrix}
1 & 0 & 0 \\
0 & 0 & 0 \\
0 & 0 & 0   
\end{pmatrix};
{H_c}=\begin{pmatrix}
0 & 1 & 0 \\
0 & 0 & 0 \\
0 & 0 & 0   
\end{pmatrix}. \label{s13}
\end{equation}
This particular $H_s$ suggests that the self-coupling link must involve the $x_{1,2}$-variable and the cross-coupling link must involve the $y_{1,2}$-variables and both be added to the dynamics of $x_1$ as suggested by $H_c$. Two ($N-1$) directed cross-coupling links (dashed arrows) are added from node 1 to nodes 2 and 3.  Replacing $H_s$, $H_c$ and using the SM model in  \eqref{s3}, the governing equations of the $i^{th}$ node is,
\begin{equation}
\begin{array}{l}
\dot{x}_i=y_i + \epsilon_1\sum\limits_{j=1}^N A_{ij} (x_j-x_i)+ \epsilon_2\sum\limits_{j=1}^NB_{ij}(y_j-y_i)\\
\dot{y}_i=r_i(x_i-x_iz_i)-by_i  \\ 
\dot{z}_i=-az_i+ax_i^2  
\end{array} \label{s14}
\end{equation}
where $i=1,2,3$ and $a=0.375, b=0.826$ and $r_i$ is the parameter used for detuning. The global stability of CS in the motif is establised in ref.\cite{SM}. Now consider $r_1=1$ for the oscillator node 1 in Fig.~\ref{fig4ab} and other two nodes 2 and 3 are perturbed so that  $r_2=2, r_3=4$, respectively. The resultant effect is exactly same as seen for two node coupled system. Attractors of the detuned (positive detuning) nodes are amplified which is reflected in the separation of three synchronization hyperplanes in Fig.~\ref{fig4ab}(b):  the  attractor of the first oscillator is plotted in a $y_1$ vs. $z_1$ plane and compared with the perturbed oscillators (2, 3) by taking a 3D plot using the $y_{2,3}$ variables along the $z$-axis.  Two of them (blue and red) are amplified with respect to the reference hyperplane (black) and their size scaling depends upon the ratios, $\frac{r_2}{r_{1}}$=2 and $\frac{r_3}{r_{1}}$=4. 
The unperturbed oscillator works as a reference here which maintains the original dynamics and accordingly, other two emerge with a GS relation with  the first. All three oscillators remain coherent, but the pertubed nodes show amplified versions of the original dynamics. The unperturbed oscillator plays a leadership role which we prove analytically (see Supplementary material \cite{SM}. 
For a better clarity of the results,  the synchronization manifolds  $y_1$ vs. $y_{2,3}$ are plotted and  the attractors of the 3-nodes are plotted in Fig.~\ref{sf3}(a) and \ref{sf3}(b), respectively. Figure \ref{sf3}(a) shows that the synchronization manifolds of all three  nodes collapse on the attractor in black when they are identical. In the case two detuned nodes (blue and red), their synchronization manifolds (blue and red lines)  are rotated away from the reference (black) where the angle of rotation is decided by $\frac{r_2}{r_{1}}$=2 and $\frac{r_3}{r_{1}}$=4, respectively.  Figure \ref{sf3}(b) shows 2D attractors (black, blue and red) where detuned nodes' attractors (blue, red) are expanded along the $y$-axis. No other variable is amplified since $x_1=x_2=x_3$ and $z_1=z_2=z_3$ is maintained. The detuned nodes' attractors expand along the $y$-axis only.

\subsection{Network motif: Ring of 3-R\"ossler Systems} 
\begin{figure}[!ht]
			\centering
			\centering
			\begin{minipage}{.2\textwidth}
				\hspace{-50pt}
				\includegraphic[width=3cm,height=3cm]{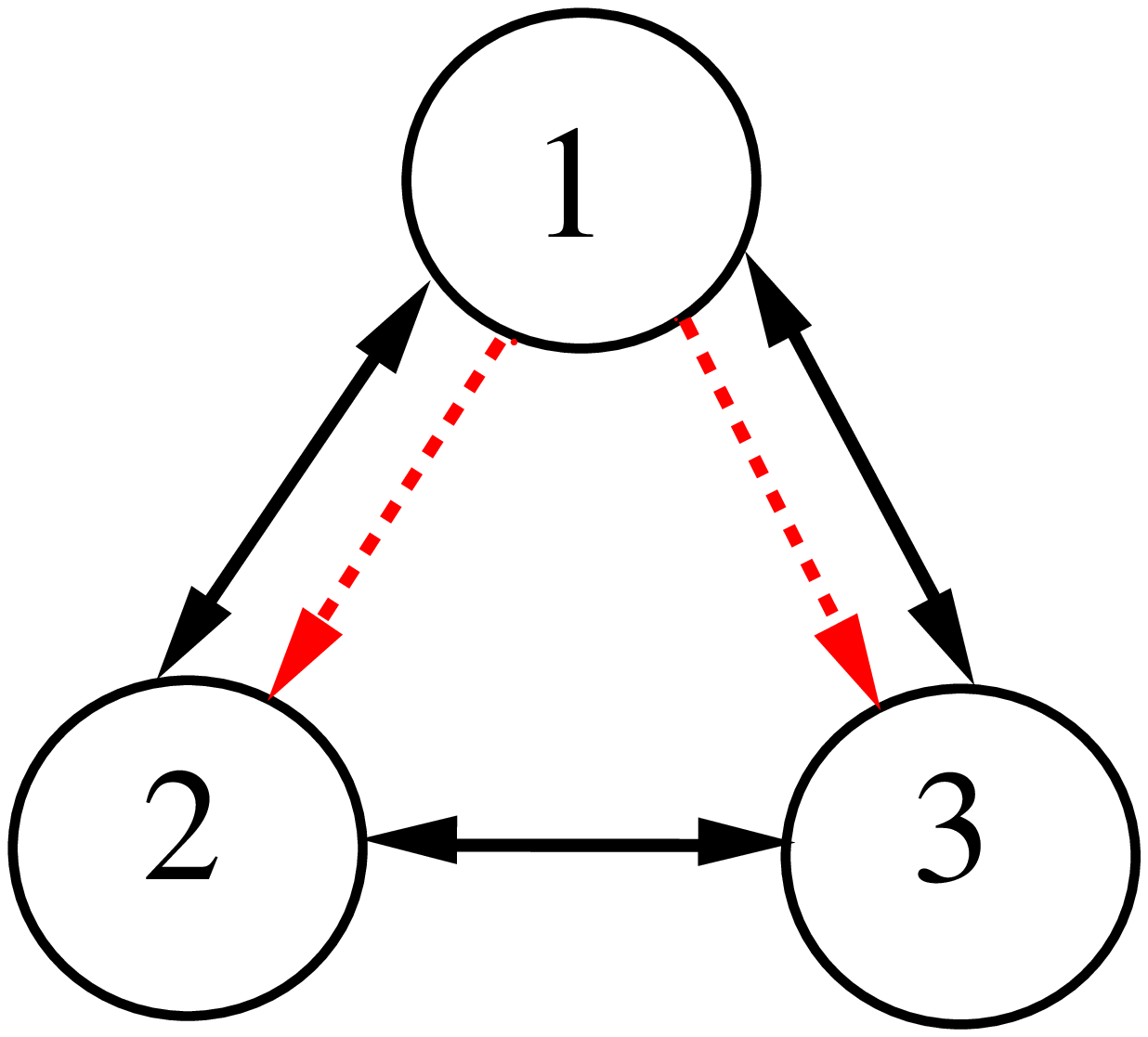}{(a)}{0pt}{0}
			\end{minipage}
			\begin{minipage}{.2\textwidth}
				\hspace{-40pt}
				\includegraphic[width=4.75cm,height=4cm]{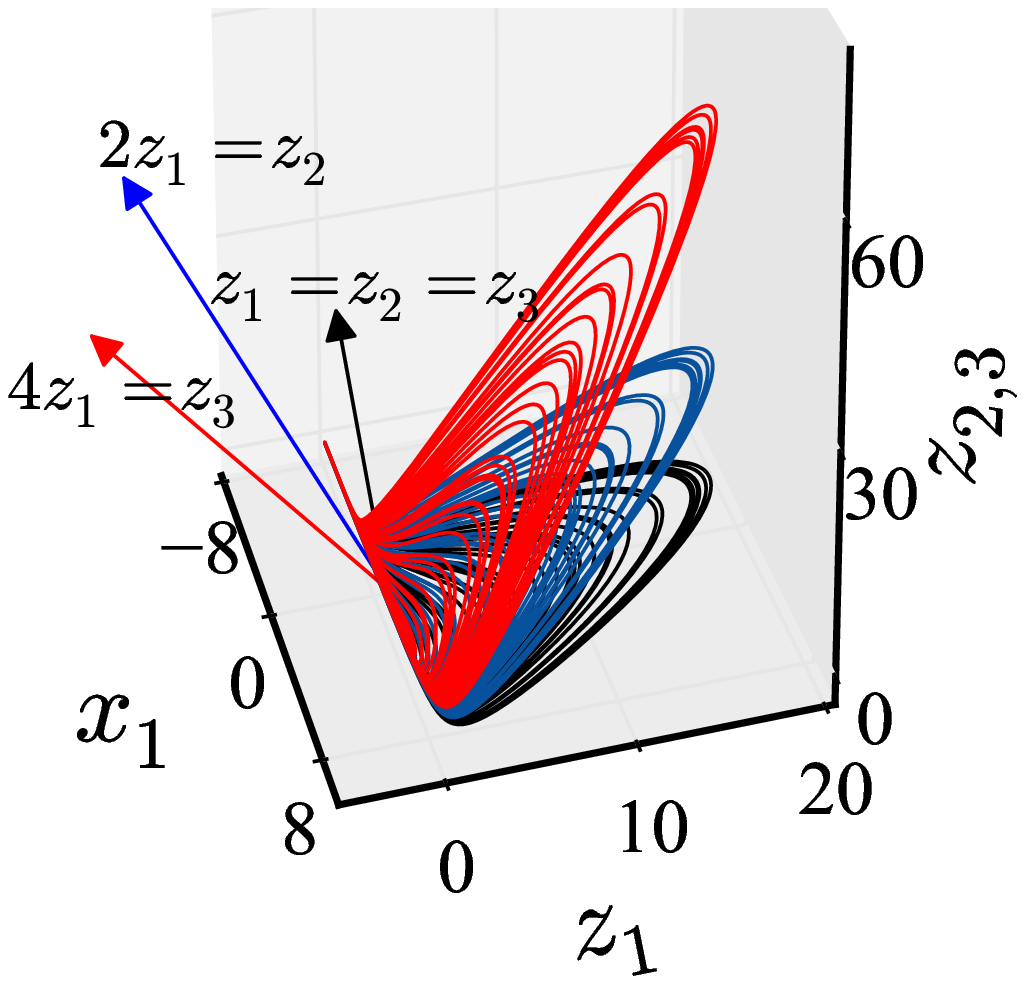}{(b)} {0pt}{0}
			\end{minipage}
\caption{(Color online) (a) Network motif: ring of 3-R\"ossler systems. (b) Projection of synchronization hyperplanes (right panel). Parameters: $a$=0.1, $c$=4.8, $b_1$=0.2, $b_2$=0.4, $b_3$=0.8 $\epsilon_1$=0.21 and $\epsilon_2$=-1.}
\vspace{-5pt}\label{fig3}
\end{figure}

	A second example of a network motif is shown in Fig.~\ref{fig3}(a), a ring of three  R\"ossler oscillators, basically, a globally coupled network. The adjacency matrix $A$ for self-coupling and the connectivity matrix of cross-coupling $B$ are
	 \[A=\begin{pmatrix}
	0 & 1 & 1 \\
	1 & 0 & 1 \\
	1 & 1 & 0
	\end{pmatrix},\;
	B=\begin{pmatrix}
	0 & 0 & 0 \\
	1 & 0 & 0 \\
	1 & 0 & 0
	\end{pmatrix}.\] 
	The coupling profile $H_s$ and $H_c$  from  \eqref{Hs} and  \eqref{Hc} for the R\"ossler system is, 
	\[ {H_s}=\begin{pmatrix}
	0 & 0 & 0 \\
	0 & 1 & 0 \\
	0 & 0 & 0
	\end{pmatrix};
	{H_c}=\begin{pmatrix}
	0 & 0 & 1 \\
	0 & 0 & 0 \\
	0 & 0 & 0
	\end{pmatrix}.\] 
	The connectivity matrix of the cross-coupling  $B$ involves 2- directed links only as shown in Fig.~\ref{fig3}(a). 		
 \begin{figure}[!ht]
	\centering
	\includegraphic[width=5cm,height=3.5cm]{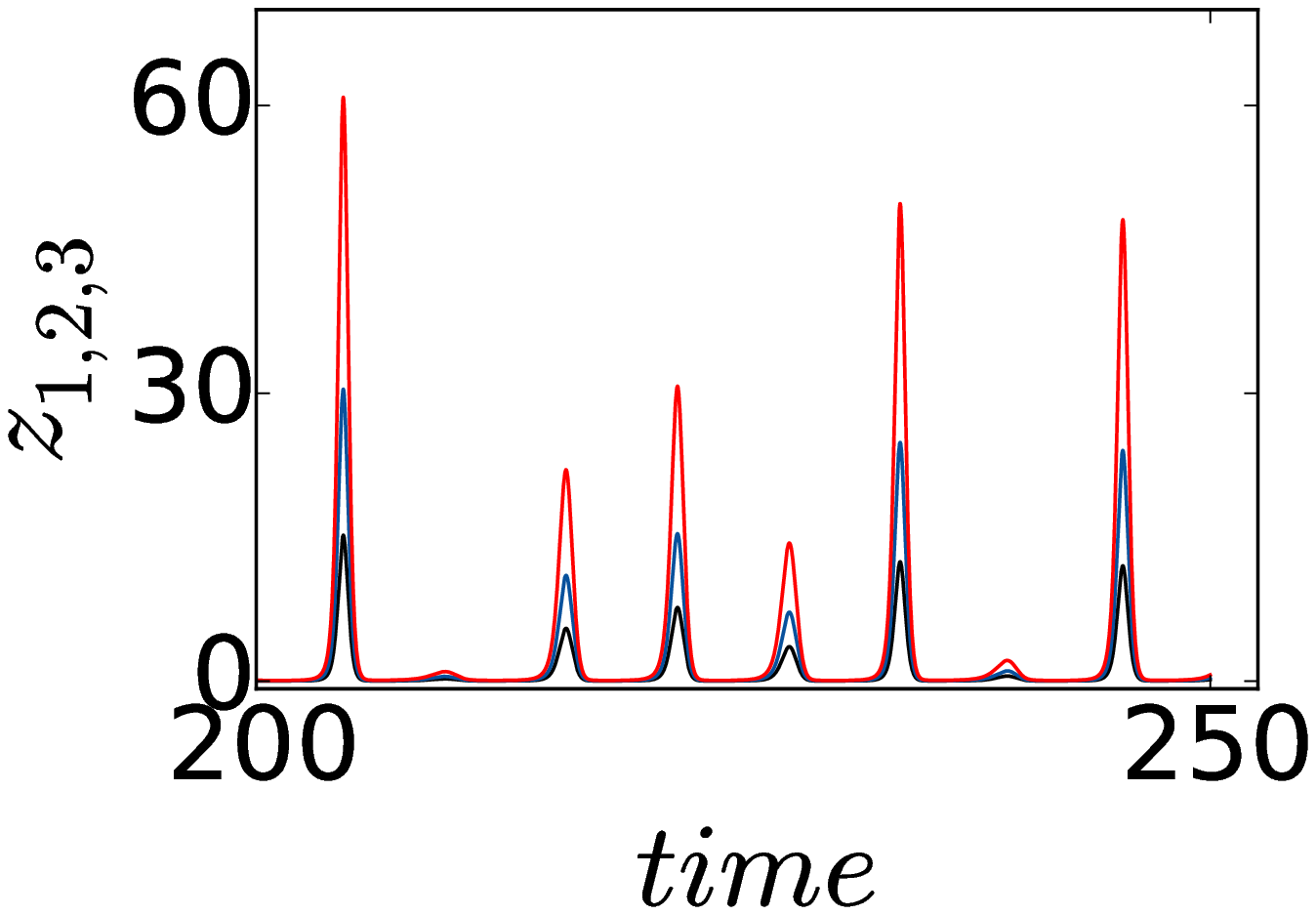}{(a)}{0pt}{0}
	\includegraphic[width=3.5cm,height=3.5cm]{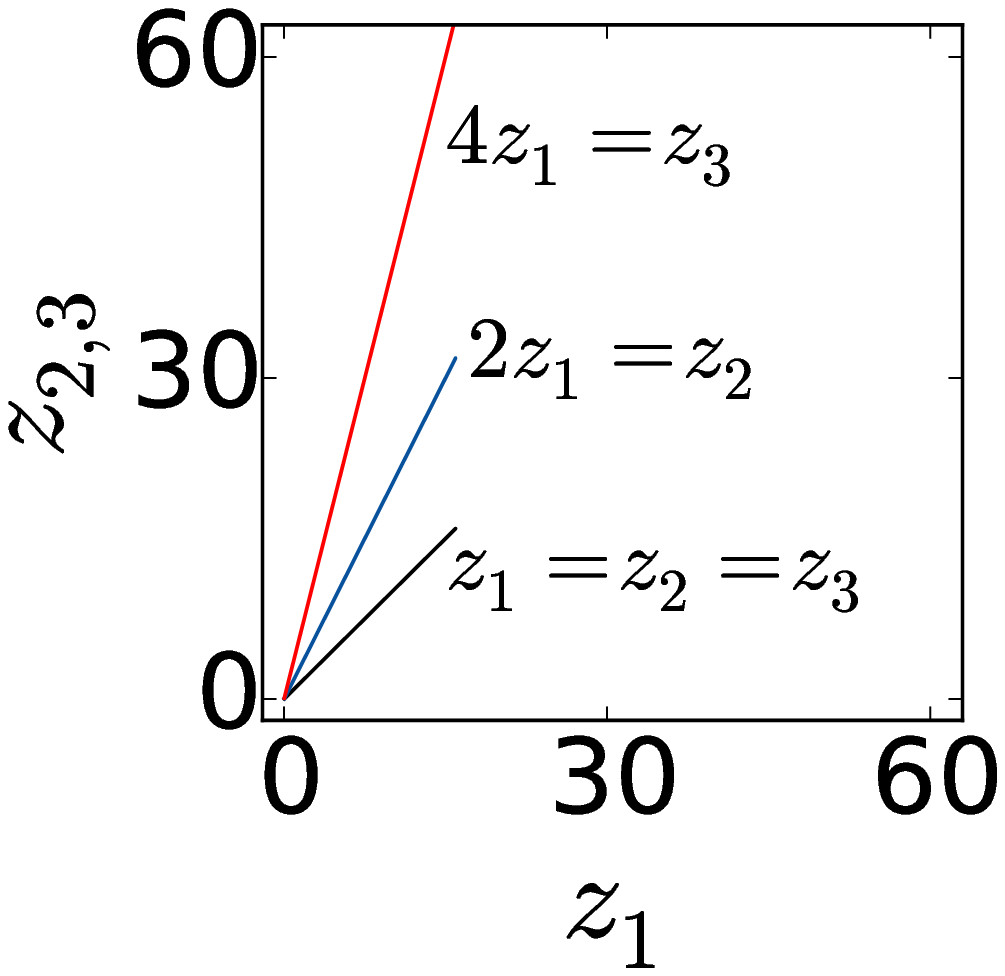}{(b)}{0pt}{0}
	\caption{(Color online) (a) Time series of $z_i$, ${i=1,2,3}$ (black, blue and red)  confirm the coherent amplification due to parameter detuning.
		(b) 2D projection of the synchronization manifolds (blue, red) are seen rotated increasingly for increasing amount of detuning, away from the reference node (black) which is the synchronization manifold for all identical oscillators.} \vspace{-5pt}\label{sf2}
\end{figure}
 The dynamics of the network-motif  under the chosen coupling profile is obtained from \eqref{s3},
\begin{equation}
\begin{array}{l}
\dot{x}_{i}=-{y}_{i}-{z}_{i} + \epsilon_2\sum\limits_{j=1}^N B_{ij}(z_j-z_i) \\
\dot{y}_{i}=x_{i}+ay_{i} +\epsilon_1\sum\limits_{j=1}^N A_{ij}(y_j-y_i)  \\ 
\dot{z}_{i}=b_i + x_{i}z_{i}-cz_{i}     
\end{array} \label{s22}
\end{equation}
$N$=3, $i\in\{1,2,3\}$.
$\textbf{x}_i=(x_i,y_i,z_i)^T$ is the state vector of the $i$-th node. The system parameters, $a = 0.2, c = 4.8$ and for identical oscillators $b_i=0.2$. The  coupling profile  realizes global stability of CS in the network-motif of identical nodes. The heterogeneity is  introduced by changing parameters ($b_1< b_2< b_3$) in two nodes of the network.  

All the nodes transits to a coherent GS state and globally stable for induced heterogeneity. In identical case ($b_1=b_2=b_3$=0.2), their dynamics will collapse to the CS hyperplane (black) as shown in Fig.~\ref{fig3}(b). For positive detuning of two nodes ($b_2=0.4, b_3=0.8$), their attractors are amplified by scaling factors $\frac{b_2}{b_1}$=2 and $\frac{b_3}{b_1}$=4 and their synchronization hyperplanes (blue, red) are simply rotated away along the $z_2$- and the $z_3$-direction, respectively, from the CS hyperplane as depicted by  their transverse direction (blue, red arrows). The attractor of the unperturbed node (black) act as the driver node ($b_1$=0.2). The amplification of the attractors is clarified by the time series plot of $z_{1,2,3}$ of three oscillators in Fig.~\ref{sf2}(a), where two of them are the amplified replica of the third one. Other variables of the detuned nodes maintain CS relation with the unperturbed node. A rotational transformation of the synchronization hyperplanes of the detuned nodes compared to the reference CS hyperplane in \ref{sf2}(b) confirms their GS relation. Note that this motif is a symmetric one due to its global structure, $B$ can be chosen arbitrarily from any one of the nodes.
\subsection{Network motif: 4-node Sprott system} \label{sec:NM4}
\begin{figure}[!ht]
		\centering
		\begin{minipage}{.2\textwidth}
			\hspace{-40pt}
			\includegraphic[width=3.5cm,height=3.5cm]{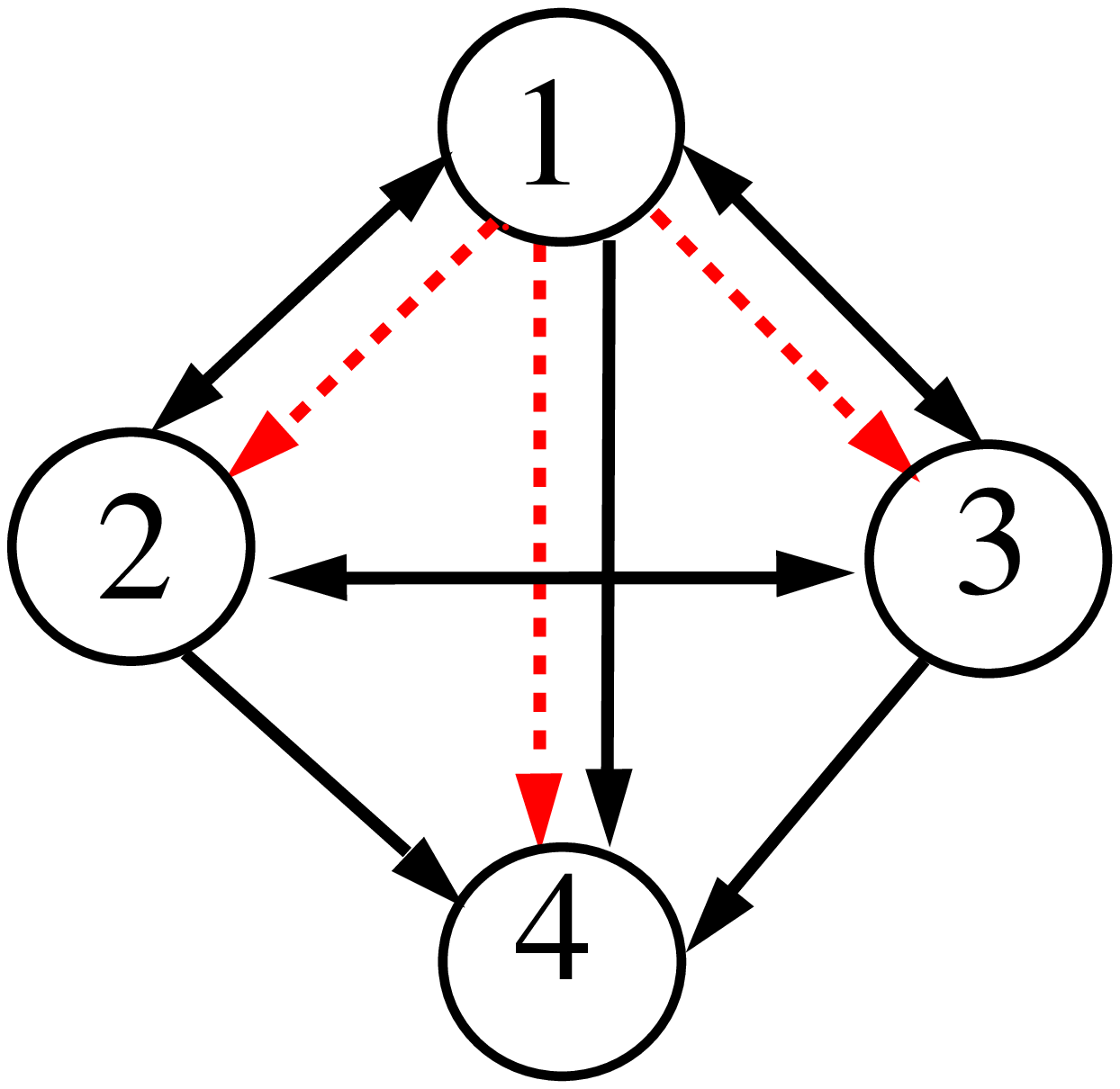}{(a)}{0pt}{0}
		\end{minipage}
		\begin{minipage}{.2\textwidth}
			\hspace{-20pt}
			\includegraphic[width=4cm,height=3.75cm]{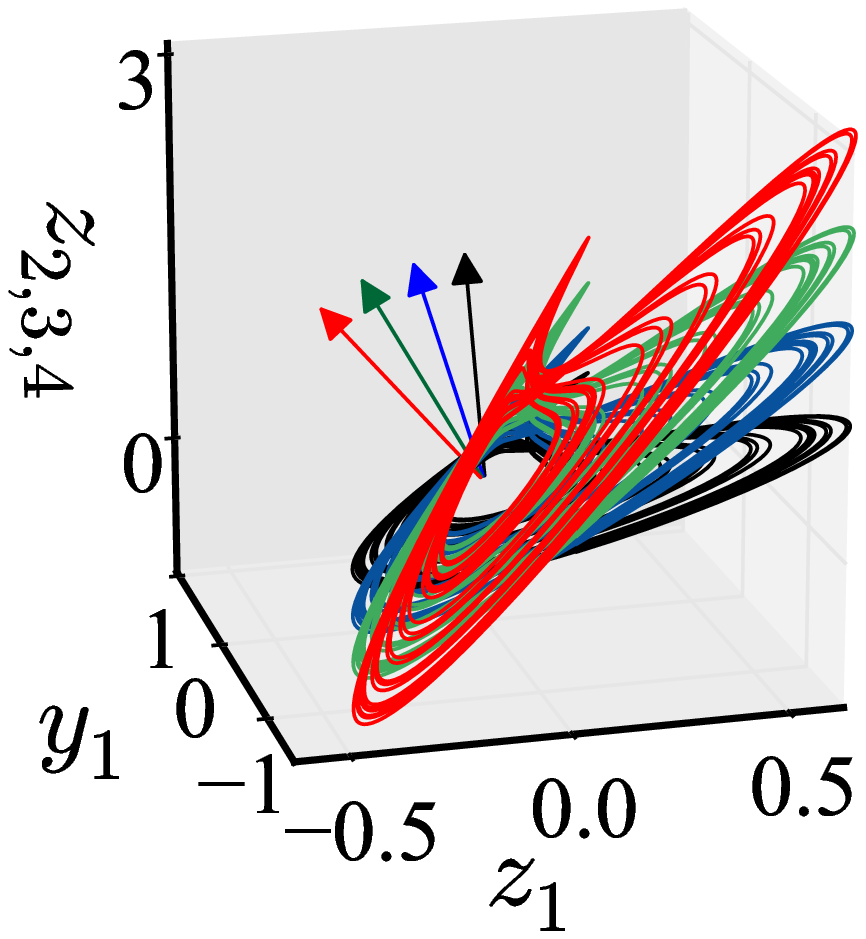}{(b)}{0pt}{0}
		\end{minipage}   	
		\caption{(Color online) (a) Schematic diagram of 4-node network motif. Solid (black) and dashed (red) arrows represent self- and cross-coupling links respectively. (b) Projection of synchronization hyperplanes. Color arrows represent rotation of transverse direction to their respective hyperplanes. Parameters are $a$=0.23, $r_1$=1, $r_2$=2, $r_3$=3, $r_4$=4, $\epsilon_1$=0.1 and $\epsilon_2$=1.}
		\vspace{-5pt} \label{fig8}
	\end{figure}

Finally,  a 4-node network-motif of the Sprott system \cite{m-5} is considered as shown in Fig.~\ref{fig8}(a). $A$ and $B$ of the 4-node network-motif are,
\begin{equation}
A=\left(
\begin{array}{cccc}
0 & 1 & 1 & 0  \\
1 & 0 & 1 & 0  \\
1 & 1 & 0 & 0  \\
1 & 1 & 1 & 0  \\
\end{array}\right), 
B=\left(
\begin{array}{cccc}
0 & 0 & 0 & 0  \\
1 & 0 & 0 & 0  \\
1 & 0 & 0 & 0  \\
1 & 0 & 0 & 0  \\
\end{array}\right)\label{s29}
\end{equation}
The coupling profile is,
	\begin{equation}
	H_s=\begin{pmatrix}
	1 & 0 & 0 \\
	0 & 0 & 0 \\
	0 & 0 & 0 
	\end{pmatrix}, \;
	H_c=\begin{pmatrix}
	0 & 0 & 0 \\
	0 & 0 & 1 \\
	0 & 0 & 0 
	\end{pmatrix}. \label{s27}
	\end{equation}
	
	Substituting them  to \eqref{s3}, the coupled dynamics of the network-motif of Sprott systems,
	\begin{equation}
	\begin{array}{l}
	\dot{x}_i=-ay_i+\epsilon_1\sum\limits_{j=1}^N A_{ij}(x_j-x_i)\\
	\dot{y}_i=x_i+z_i+\epsilon_2\sum\limits_{j=1}^N B_{ij}(z_j-z_i)\\
	\dot{z}_i=r_i(x_i+y_i^2)-z_i,
	\end{array} \label{s28}
	\end{equation}
	where $a$=0.23 and $r_i$ is the heterogeneity parameter.	
	As shown in Fig.~\ref{fig8}(b), the projected synchronous hyperplanes for three perturbed  nodes are rotated away  (blue, green and red, respectively, along the $z_2$-, $z_3$- and $z_4$-axis) from the reference hyperplane (black) and their respective scaling factors are $\frac{r_{i}}{r_1}$, $i=2,3,4$.  
	\begin{figure}[!ht]
		\centering
		\includegraphic[width=4.25cm,height=3.25cm]{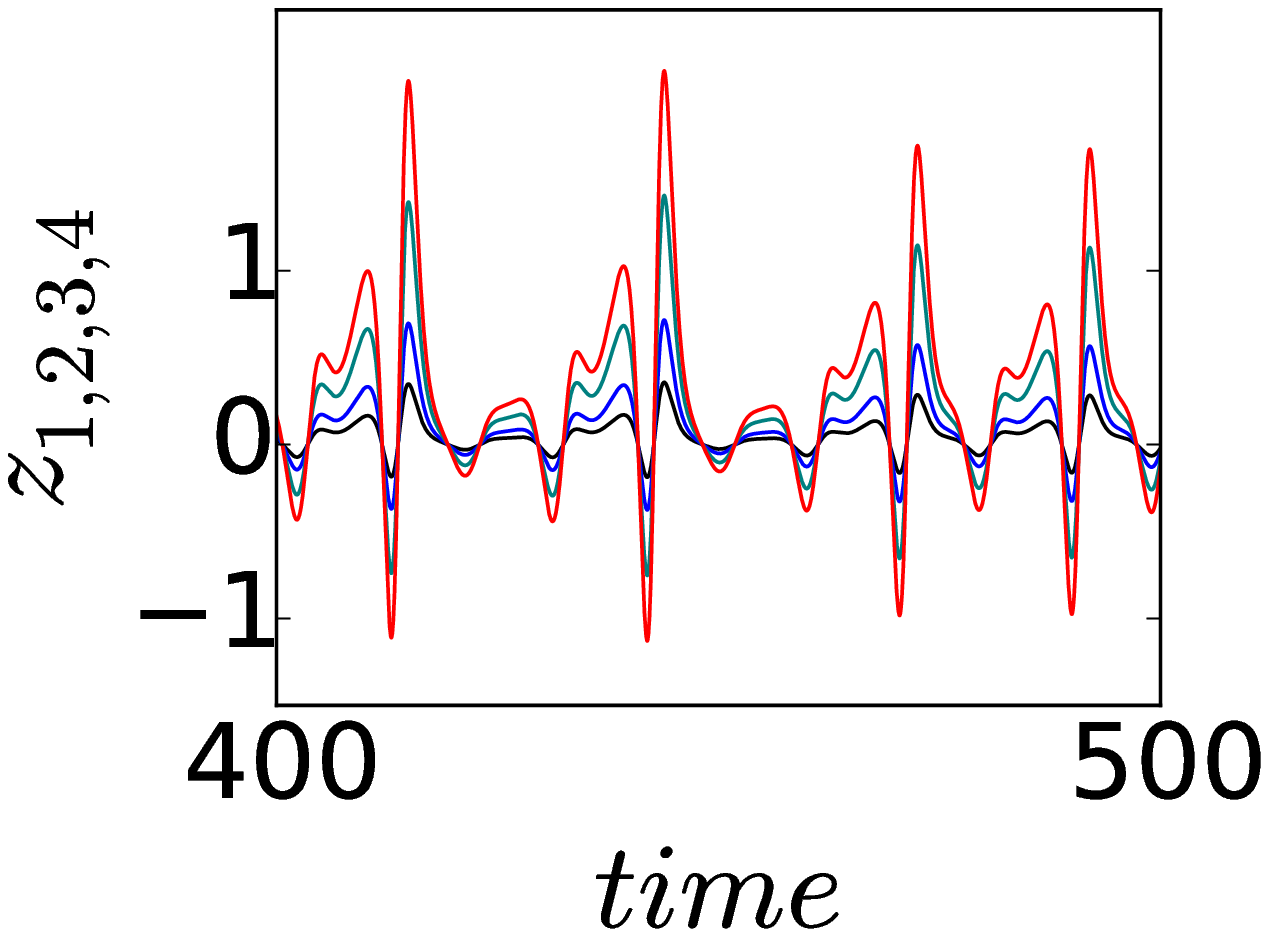}{(a)}{0pt}{0}
		\includegraphic[width=4.25cm,height=3.25cm]{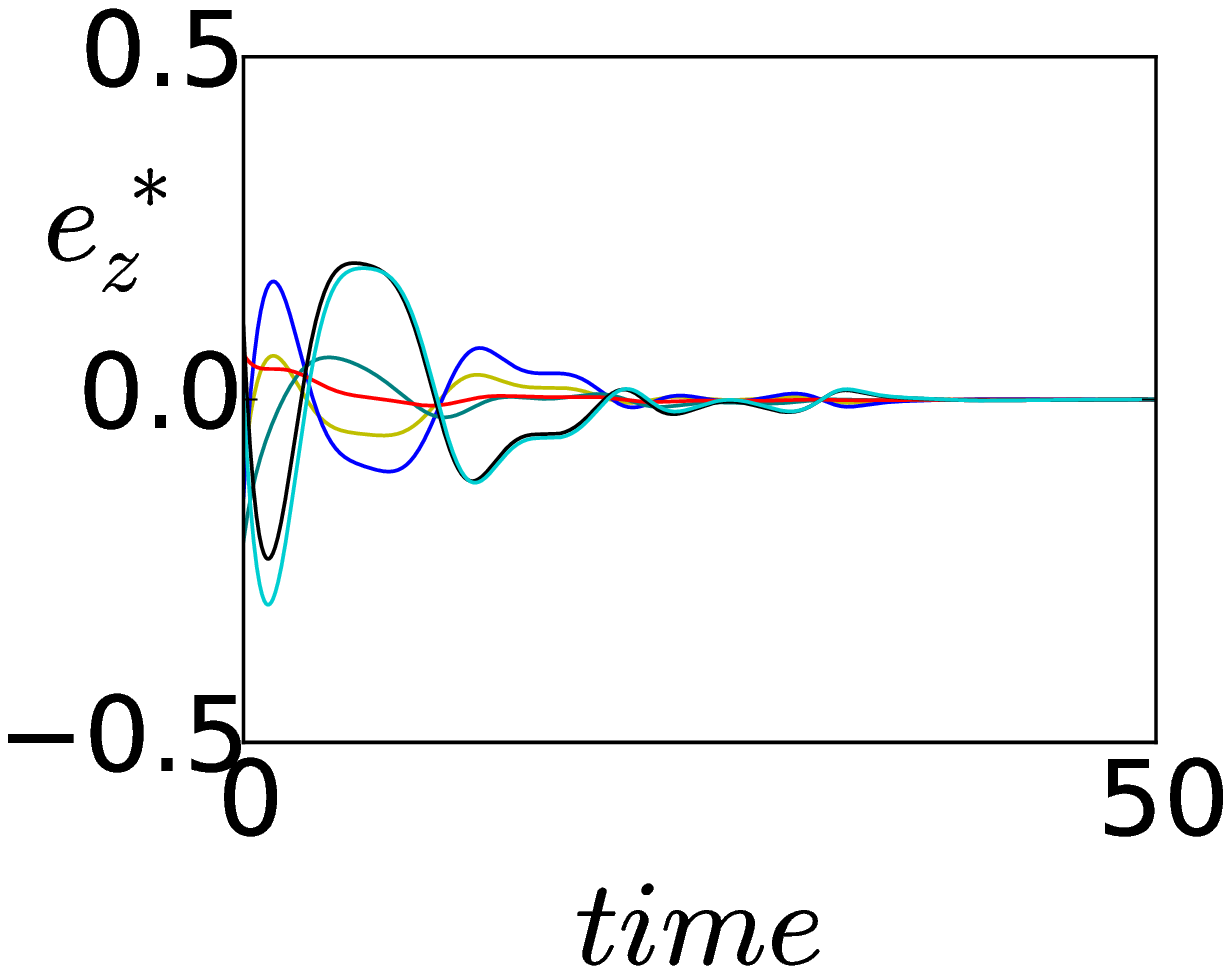}{(b)}{0pt}{0}
		\caption{(Color online) (a) Time series of $z_{1,2,3,4}$ of four oscillators, which confirms amplification. (b) Modified error $e_z^*$, which converges to zero as $t\rightarrow\infty$. }		\vspace{-5pt}\label{sf4}
	\end{figure}
	Further numerical confirmation is given in  Fig.~\ref{sf4}(a) where it is  clearly seen that the time series of $z_{1,2,3,4}$-variables are affected by the induced heterogeneity of three nodes. Actually, three of them are amplified but remain coherent  with all each other. This is confirmed by the modified error plot $e^*$ in Fig.~(Fig.~\ref{sf4}(b)) that converges to zero in time when $z_1=\frac{z_2}{2}=\frac{z_3}{3}=\frac{z_4}{4}$. Other variables remain unchanged and maintain CS relation in all three nodes. 
\section{Constructive role of cross-coupling in a network}
\begin{figure}[!ht]
	\centering
	\includegraphics[width=6cm,height=6cm]{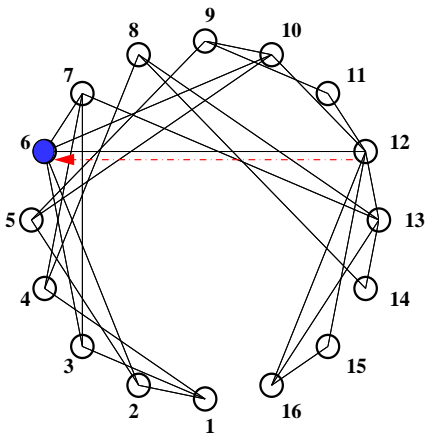}
	\caption{(Color online) Schematic diagram of a network. Node-6 (blue node) is perturbed while all other nodes have identical parameters. To restore synchrony only one directed cross-coupling link (dashed red arrow) is added to the node-6 from a neighbor node-12.}\label{sens2}
\end{figure}

\begin{figure}[!ht]
	\centering
	\includegraphic[width=6cm,height=4cm]{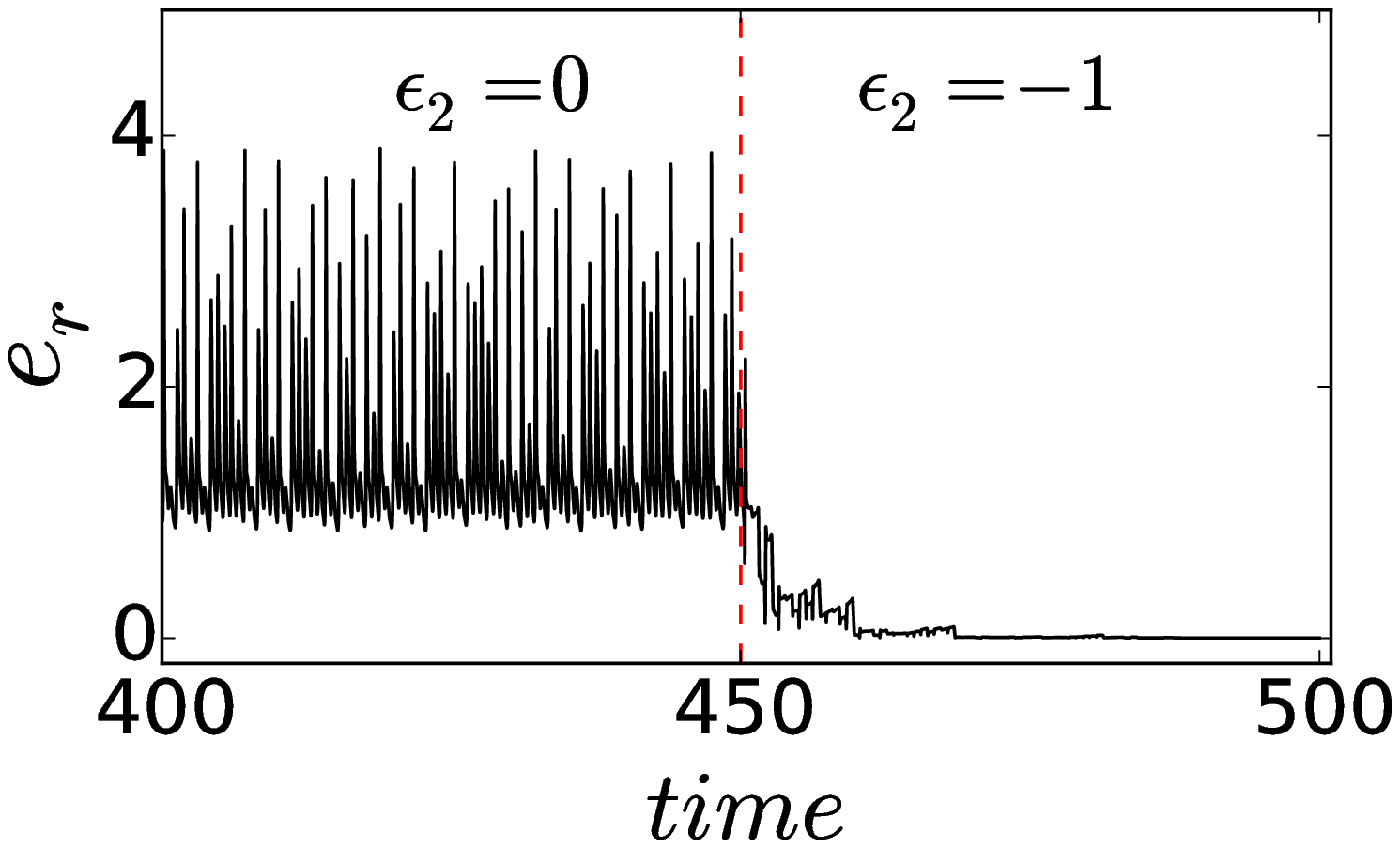}{(a)}{0pt}{0}
	\includegraphic[width=4.65cm,height=4.35cm]{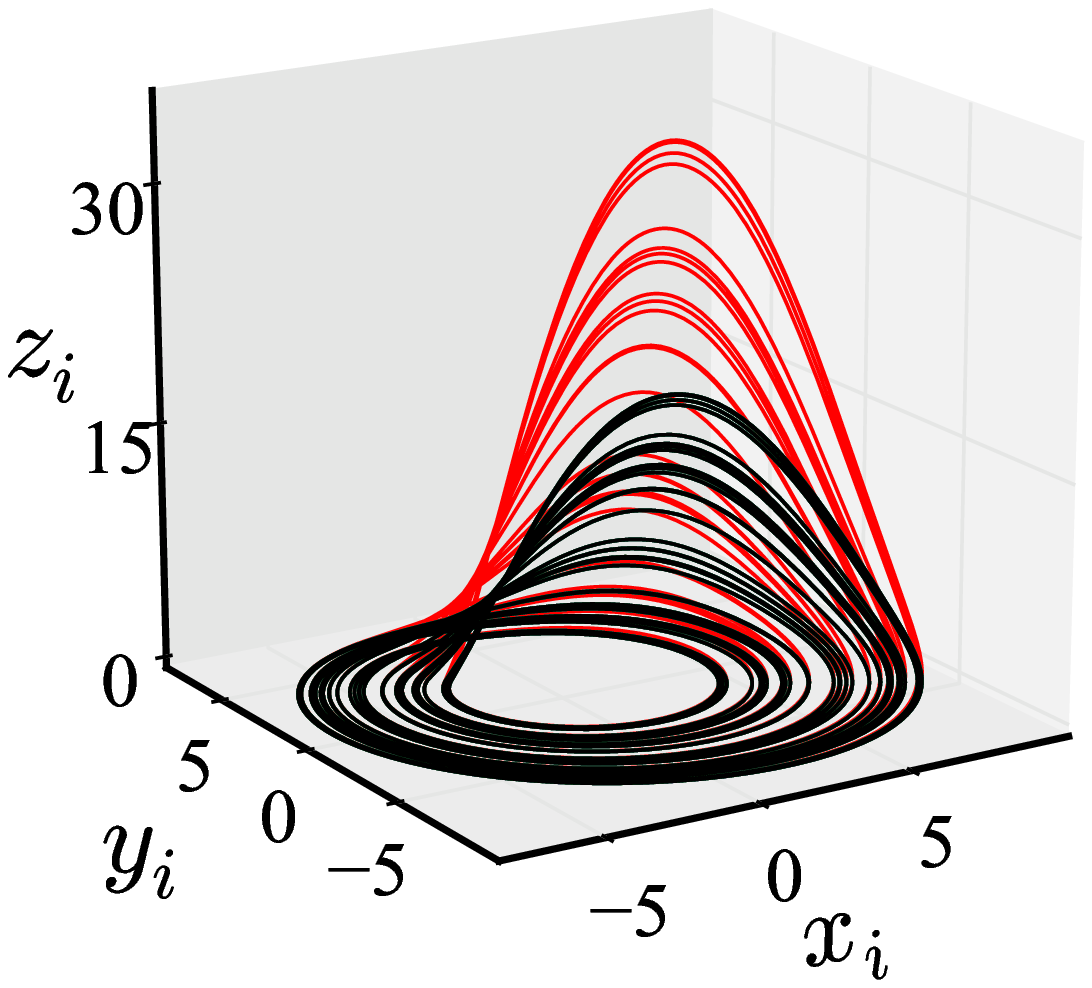}{(b)}{0pt}{0}
	\caption{(Color online) Network of coupled R\"ossler system. $a=0.2,c=4.8$ and $b_i=b=0.2$, where $i$=$1,...,N$ with self-coupling strength $\epsilon_1=0.21$.  (a) Time series of the synchronization error $e_r$ shows that in absence of the cross-coupling link ($\epsilon_2=0$), synchrony is lost when node-6 is perturbed ($b\neq b_6=0.4$). At ($t\geq 450$)  one directed cross-coupling link ($\epsilon_2=-1$) is adding from node-12 to node-6, synchrony is restored ($e_r=0$) in the network. (b) After $t=450$, attractors of node-6 and rest of the nodes are shown in gray (red) and black lines, respectively, which confirm amplification along the $z$-variable of node-6 compared to attractors of other nodes.} \label{sens2_result}
\end{figure}

It is demonstrated here how a single directed cross-coupling link makes a dramatic change in the quality of synchronization in a  larger network shown in Fig.~\ref{sens2}.  The self-coupling links define the network topology (adjacency matrix $A$)  in solid lines whose each node is represented by a R\"ossler system. The self-coupling matrix between any two nodes is  chosen from the LFM ($H_s$ in Eq.~\eqref{Hs} for R\"ossler system) to realize CS for all identical nodes (strength of each self-coupling link, $\epsilon_{1}$=0.21, $\epsilon_2$=0 in Eq.\eqref{s22}). When node-6 (blue circle) is perturbed, synchrony is lost as shown in Fig.~\ref{sens2_result}(a). One directed cross-coupling link ($\epsilon_2=-1$) as defined by $H_c$ in Eqn.(10) for a Rossler system is now added from node-12 to node-6 (dashed red arrow). Numerically solving Eq.~\eqref{s22} for N=16, we find that synchrony is restored in the network as shown in Fig.~\ref{sens2_result}(a) after an instant $t=450$ when the cross-coupling link is added. Surprisinly, all the nodes return to CS and follows the isolated dynamics except the node-6 whose attractor is simply amplified/attenuated for a positive/negative detuning as shown in Fig.~\ref{sens2_result}(b). Effectively, node-12 is now playing the role of a driver to node-6 as discussed above for network-motifs. In fact, one cross-coupling link from any one of the neighboring nodes (nodes, 2, 3, 7, 10, 12) to the perturbed node-6 will produce the same result. Synchronization in the network is monitored by defining the error $e_r$ as $e_r(t)=\sqrt{\sum_{i=1}^{N}\left[(\tilde{x}-x_i)^2+(\tilde{y}-y_i)^2\right]}$, where $N$ is the network size,  $\tilde{x}=\frac{1}{N}\sum_{i=1}^{N}x_i$ and $\tilde{y}=\frac{1}{N}\sum_{i=1}^{N}y_i$. $e_r\neq 0$ means no synchronization.

\section{Conclusion}
\par We proposed a set of general coupling conditions for a  selection of a coupling profile from the LFM of a dynamical system that realized globally stable synchrony  in an ensemble of the system and a robustness of synchrony to perturbation in system parameter as well as the coupling parameter.  
 The coupling profile consisted of self-coupling  and cross-coupling matrices that defined the coupling links between any two nodes of a network. We explained how the selection of the self-coupling  and the cross-coupling matrices were systematically made from the nature of the diagonal  and  off-diagonal elements, respectively, of the LFM of a dynamical system.  
Addition of selective cross-coupling links over and above the conventional diffusive coupling (defined here as self-coupling) in particular made a dramatic improvement in the quality of synchronization. Besides establising global stability of CS in the coupled system, it expands the range of critical coupling for CS from the range defined by the conventional MSF conditions and as a result, it saves synchrony from a breakdown in a situation of a drifting of the coupling parameter. Most importantly, the addition of the cross-coupling link added robustness to synchrony against large mismatch of system parameters. We demonstrated the  constructive roles of our proposed coupling scheme, especially, the cross-coupling link using many example systems. We exemplified the  constuctive role using an example of a larger  network in a CS state by a manual detuning of a system parameter  when all the nodes of the network remained unpertubed in the CS state, but  the perturbd node showed a transition from CS to GS. The manifestation of the GS state is  an amplified (attenuated) response of the perturbed node`s attractor for a positive  (negative) detuning of the system parameter. We  validated our proposed coupling profile scheme with examples of 2- coupled systems, 3-node and 4-node network motifs  using the Lorenz system, the R\"ossler system, the Hindmarsh Rose model, the Shimizu-Morioka laser model and a Sprott system as dynamical nodes. We  supported our results by analytical as well as numerical verifications in all the example systems except the large network. The coupling scheme worked efficiently for many other network-motifs (see Supplementary Material). We plan to work for more rigiorus analysis of our results for large networks.\\
\begin{acknowledgments}
  P.K.R. acknowledge support by the CSIR (India) Emeritus scientist scheme. A.M. is supported by the UGC (India). S.K.D. is supported by the University Grants Commission (India) Emeritus Fellowsip. S.K.B. and S.S. acknowledge support by the DST-SERB (India) File No.-YSS/2014/000687.
\end{acknowledgments}
\section*{Appendix: Two coupled systems}
{\bf Shimizu-Morioka model:}
From \eqref{Hs} and \eqref{Hc}, bidirectional self-couplings are introduced through the $x_{1,2}$ variables and one cross-coupling link involves $y_{1,2}$ variables and added to the dynamics of $x_1$ so that the governing equations are,
\begin{equation}
\begin{array}{l}
\dot{x}_1=y_1 + \epsilon_1 (x_2-x_1)+ \epsilon_2(y_2-y_1)\\
\dot{x}_2=y_2 + \epsilon_1 (x_2-x_1) \\
\dot{y}_{1,2}=r_{1,2}(x_{1,2}-x_{1,2}z_{1,2})-by_{1,2} \\ 
\dot{z}_{1,2}=-az_{1,2}+ax_{1,2}^2  
\end{array} \label{s114}
\end{equation}
Error dynamics are 
\begin{equation}
\begin{array}{l}
\dot{e_x}={e_y}+\epsilon_1(-{e_x}-{e_x})-\epsilon_2{e_y}
=-2\epsilon_1{e_x}+(1-\epsilon_2){e_y} \\
\dot{e_y}=r_1(x_1-x_1z_1)-r_2(x_2-x_2z_2)-b{e_y}\\
\dot{e_z}= -a{e_z}+a(x_1^2-x_2^2)\\
\end{array}\label{s116}
\end{equation}
We first examine the synchronization of $x_1=x_2$ by  considering a Lyapunov function $V(e_x)=1/2e_x^2$. The time derivative of $V(e_x$) is,
\begin{equation}
\dot{V}({e}_x)={e}_x\dot{{e}}_x=-2\epsilon_1{e_x^2}+(1-\epsilon_2)({e_x}{e_y}) \label{s117}
\end{equation}
\begin{equation}
 \quad \dot{V}({e}_x)=-2\epsilon_1{e_x}^2, \label{s118}
\end{equation}
if  ${\epsilon_2}$=1 and $\epsilon_1>0;$ 
when $\lim_{t\rightarrow\infty} {e}_x(t) \rightarrow 0,$ 
that implies $x_1=x_2$ is asymptotically stable. Using this condition,  we check the stability of $z_1=z_2$ when we find, 
\begin{equation}
\dot{V}(e_z)=e_z\dot{e}_z= -a{e_z}^2<0. \label{s119}
\end{equation}
$z_1=z_2$ is now asymptotically stable as $t\rightarrow\infty$. This implies $\dot e_y$=-$e_y$ for identical oscillators, $r_1=r_2$. The global stability of CS ($x_1=x_2$, $y_1=y_2$ and $z_1=z_2$) is thus established. \\
For a parameter detuning of $r_1$ of oscillator-1 from its identical values ($r_1\neq r_2$), the error function $e_y$ conly changes, \\
$\dot{e_y} =(r_1-r_2) (x_1-x_1z_1)-b{e_y} = \frac{r_1-r_2}{r_1}(\dot{y}_1+by_1)-b{e_y}$;\\
this implies $\dot{y}_{1}\frac{r_2}{r_1}-\dot{y}_2=-b(\frac{r_2}{r_1}y_1-y_2)$, or, $\dot{e_y^*}=-b{e_y^*}$, where the error variable is redefined by ${e_y^*}=y_1 \frac{r_2}{r_1}-y_2$. \\
One can obtain the time derivative of the  Lyapunov function for V($e^*_y$) as
\begin{equation}
\dot{V^{'}}(e_y^*)=e_y^*\dot{e}_y^*=-b{e_y^{*2}}. \label{s120}
\end{equation} 
Therefore, under the conditions  ${\epsilon_2}=1$ and $\epsilon_1>0$, a new coherent state, ($x_1=x_2$, $y_1=\frac{y_1}{y_2}$ and $z_1=z_2$) emerges which we call as a GS state and it is globally stable. \\\\
{\bf R\"ossler system:} 
From \eqref{Hs} and \eqref{Hc} the coupled R\"ossler system, the coupling profile appears as a bidirectional self-coupling that involves $y_{1,2}$ variables. The cross-coupling involves $z_{1,2}$ variables and added the dynamics of $x_1$. The dynamical equations of two coupled R\"ossler systems, 
\begin{equation}
\begin{array}{l}
\dot{x}_{1}=-{y}_{1}-{z}_{1} + \epsilon_2(z_2-z_1) \\
\dot{x}_{2}=-{y}_{2}-{z}_{2}  \\
\dot{y}_{1,2}=x_{1,2}+ay_{1,2} +\epsilon_1(y_{2,1}-y_{1,2})  \\ 
\dot{z}_{1,2}=b_{1,2} + x_{1,2}z_{1,2}-cz_{1,2}     
\end{array} \label{s122}
\end{equation}
 The error dynamics is
\begin{equation}
\begin{array}{l}
\dot{{e_x}}=-{e_y}-{e_z} - \epsilon_2{e_z}\\
\dot{{e_y}}={e_x}+(a-2\epsilon_1){e_y}\\
\dot{e_z}=b_1 +x_1z_1 -b_2+x_2z_2- c{e_z} \\ \label{s123}
\end{array}
\end{equation}	
We determine the stability of $x_1=x_2$ and $y_1=y_2$ first. This involves the construction of a Lyapunov function $V(e_x,e_y)$ which is positive definite function, $V(e_x,e_y)=\frac{1}{2}(e_x^2+e_y^2).$ 
The time derivative of the Lyapunov function,
\begin{align}
\dot{V}(e_x,e_y) =e_x\dot{e}_x+e_y\dot{e}_y
=-(1+ \epsilon_2)e_xe_z+(a-2 \epsilon_1)e_y^2.\label{s124}
\end{align}	
For ${\epsilon_1}>a/2$ and ${\epsilon_{2}}=-1$, $\dot{V}(e_x,e_y)\leq0$ is negative semidefinite, since we get $\dot{V}(e_x,e_y) = 0$ if $e_y = 0$ and for any values of $e_x$.  However, by using the LaSalle invariance principle \cite{LaSalle}, the set $S=\{e_x,e_y\}$ does not contain any trajectory except the trivial trajectory $(e_x,e_y)$=0, as $\dot{e_y}\neq0$ if $e_x\neq0$. As a result, the trajectory will not stay in the set $S$. So the partial synchronization manifold of $x_1=x_2$ and $y_1=y_2$ is asymptotically stable with  $t\rightarrow\infty$ above a critical value of $\epsilon_1$. We get an unbounded synchronization region for ${\epsilon_1}_c>a/2$ which is interestingly lower than the critical value found by linear stability analysis (MSF) in \cite{Olusola}. As a result, the threshold region of self-coupling is expanded by the addition of the cross-coupling link. We consider the error dynamics $\dot e_z$ separately when considering the effect of heterogeneity.
\par  Assume oscillators have different parameters $b_1\neq b_2$.
The stability in $e_x$ and $e_y$ is undisturbed by the induced heterogeneity when $\dot{V}(e_x,e_y)\leq0$ remains valid. 
Now we check the error dynamics ${e_z}$,
\begin{align*}
&\dot{e_z}=(b_1-b_2) + x_1e_z - ce_z \\
&=\frac{b_1-b_2}b_1\left( \dot{z}_1-x_1z_1+cz_1\right)+ x_1e_z - ce_z\\
&\dot{e_z}-\frac{b_1-b_2}{b_1}\dot{z}_1 =x_1({e_z}-\frac{b_1-b_2}{b_1}z_1)-c(e_z-\frac{b_1-b_2}{b_1}z_1) \\
&\dot{e_z^*} = -ce_z^*(1-x_1/c),
\text{ where } {e^*_z}=z_1\frac{b_2}{b_1}-z_2.
\end{align*}
From the modified error dynamics of $e_z$ one can write $\dot{V^{'}}(e_z^*)=e_z^{*}\dot{e}_z^* = -ce_z^{*2}(1-x_1/c)$.
Hence, using the LaSalle invariance principle \cite{LaSalle} the coupled R\"ossler system \eqref{s122} is globally synchronized provided $c>\mid x_1\mid$. In fact, it shows an error to the limit of $10^{-4}$. Note that this asymptotic stability of $z_1=z_2$ is valid even for identical systems when $\dot V(e_z)<0$.\\\\
{\bf Sprott system:}
We provide analytical evidence for the stability of two coupled Sprott systems. According to \eqref{Hs} and \eqref{Hc}, the coupling profile consists of a bidirectional self-coupling involving $x_{1,2}$ variables and a cross-coupling involving $z_{1,2}$ variables is added to $y_1$ dynamics. The coupled Sprott system is,
\begin{equation}
\begin{array}{l}
\dot{x}_{1,2}=-ay_{1,2}+\epsilon_1(x_{2,1}-x_{1,2})\\
\dot{y}_1=x_1+z_1+\epsilon_2(z_2-z_1)\\
\dot{y}_2=x_2+z_2\\		
\dot{z}_{1,2}=r_{1,2}(x_{1,2}+y_{1,2}^2)-z_{1,2},
\end{array} \label{s128}
\end{equation}
Error dynamics of the system \eqref{s128},
\begin{equation}
\begin{array}{l}
\dot{{e}_x}= -a{e_y}-2\epsilon_1{e_x} \\
\dot{{e}_y}= e_x+(1-\epsilon_2)e_z \\
\dot{e_z}=r_1(x_1+y_1^2)-r_2(x_2+y_2^2)-e_z 
\end{array} \label{s130}
\end{equation}
To prove global stability, a Lyapunov function is constructed, $V(e_x,e_y,e_z)=\frac{1}{2a}e_x^2+\frac{1}{2}e_y^2+\frac{1}{2}e_z^2.$ 
From \eqref{s130}, we examine synchronization in $x$ and $y$ variables of the system \eqref{s128}. We check the stability of $e_x$=$e_y$=0 first. Now the Lyapunov function is defined as $V(e_x,e_y)=\frac{1}{2a}e_x^2+\frac{1}{2}e_y^2$. The time derivative of this Lyapunov function is
\begin{align}
\dot{V}(e_x,e_y)=\frac{1}{a}e_x\dot{e}_x+e_y\dot{e}_y 
=&-\frac{2\epsilon_1}{a}e_x^2 +(1-\epsilon_2)e_ye_z. \label{s133}
\end{align} 
The criterion for Eq. \eqref{s130} to be negative semidefinite, i.e. $\dot{V}(e_x,e_y) \leq 0$ are:
\begin{equation}
\epsilon_2=1  \quad \text{and} \quad \epsilon_1>0. \label{s134}
\end{equation}
Under the criterion \eqref{s134} and using the LaSalle invariance principle \cite{LaSalle}, $x_1=x_2$ and $y_1=y_2$ state is stable as $t\rightarrow  \infty $.
Using the above partial synchronization condition in \eqref{s130} and after a detuning of $r_{1,2}$ parameter, we obtain modified error dynamics,
\begin{align}
&\dot{e_z} = (r_1-r_2) (x_1+y_1^2)-e_z = \frac{r_1-r_2}{r_1} (\dot{z_1}+z_1)-e_z  \nonumber \\
&\dot{e_z}- \frac{r_1-r_2}{r_1}\dot{z_1}=  \frac{r_1-r_2}{r_1} z_1-{e_z}  \nonumber  \\
&\dot{e_z^*} =-e_z^* 
\label{s140}
\end{align}
where the modified error variable is,
\begin{equation}
\begin{array}{l}
{e_z^*}=z_1\frac{r_2}{r_1}-z_2		
\end{array}
\end{equation}
and when the time derivative of the Lyapunov function, 
\begin{align}
\dot{V}^{'}(e_z^*)&=e_z^*\dot{e}_z^* = -e_z^{*2} < 0. \label{s141}
\end{align}
Applying the LaSalle invariance principle \cite{LaSalle} once again and under the condition \eqref{s134}, the system \eqref{s128} is globally synchronized in a GS state defined by $x_1=x_2$, $y_1=y_2$ and $z_1=\frac{r_1}{r_2}z_2$.


\begin{thebibliography}{100}
\bibitem{equ1} $\dot{e}_y= r_1x_1-r_2x_2-e_y-x_1z_1+x_2z_2$. As, $x_1=x_2$; $\dot{e}_y= \Delta r x_1-e_y-x_1e_z$ = $-\frac{\Delta r}{r_1} (\dot{y}_1+y_1+x_1z_1)-e_y-x_1e_z$; $\dot{y}_1+\frac{\Delta r}{r_1}\dot{y}_1-\dot{y}_2=-\frac{\Delta r}{r_1}y_1-e_y+x_1(\frac{\Delta r}{r_1}z_1-e_z)$. Now, $\dot{e}_y^*=-e_y^*-x_1e_z^*$. Where the modified error variables are, $e_y^*=\frac{\Delta r}{r_1}y_1-y_2$ and $e_z^*=\frac{\Delta r}{r_1}z_1-z_2$.
\end{thebibliography}
\end{document}